\documentclass[12pt]{article}
\usepackage{amsmath, amsthm, amssymb}
\usepackage{setspace}
\usepackage{multirow}
\usepackage{graphicx}
\usepackage{apacite}
\usepackage{natbib}
\usepackage{color}
\usepackage{cite}
\usepackage{hyperref}

\newcommand{\uD}       {\mbox{\boldmath$D$}}
\newcommand{\ud}       {\mbox{\boldmath$d$}}

\newcommand{\uI}       {\mbox{\boldmath$I$}}

\newcommand{\uu}       {\mbox{\boldmath$u$}}

\newcommand{\uw}       {\mbox{\boldmath$w$}}
\newcommand{\uX}       {\mbox{\boldmath$X$}}
\newcommand{\ux}       {\mbox{\boldmath$x$}}
\newcommand{\uY}       {\mbox{\boldmath$Y$}}
\newcommand{\uy}       {\mbox{\boldmath$y$}}
\newcommand{\uZ}       {\mbox{\boldmath$Z$}}
\newcommand{\uz}       {\mbox{\boldmath$z$}}

\newcommand{\ualpha}       {\mbox{\boldmath$\alpha$}}
\newcommand{\ubeta}       {\mbox{\boldmath$\beta$}}
\newcommand{\ugamma}       {\mbox{\boldmath$\gamma$}}
\newcommand{\uDelta}       {\mbox{\boldmath$\Delta$}}
\newcommand{\udelta}       {\mbox{\boldmath$\delta$}}
\newcommand{\uepsilon}       {\mbox{\boldmath$\epsilon$}}

\newcommand{\umu}       {\mbox{\boldmath$\mu$}}

\newcommand{\uxi}       {\mbox{\boldmath$\xi$}}

\newcommand{\uSigma}       {\mbox{\boldmath$\Sigma$}}
\newcommand{\usigma}       {\mbox{\boldmath$\sigma$}}

\newcommand{\uzero}       {\mbox{\boldmath$0$}}
\newcommand{\uone}       {\mbox{\boldmath$1$}}

\newcommand{\ahat} {\hat{\boldsymbol{\hspace{-2pt}\alpha}}}

\newcommand{\utX} {\tilde{\boldsymbol{\hspace{-2pt}X}}}

\newcommand{\uty} {\tilde{\boldsymbol{\hspace{-2pt}y}}}

\newcommand{\pkg}[1]{\textbf{#1}}
\newcommand{\CRANpkg}[1]{\href{https://CRAN.R-project.org/package=#1}{\pkg{#1}}}
\newcommand{\code}{\texttt}

\title{BNSP: an R Package for Fitting Bayesian Semiparametric Regression Models and Variable Selection}

\author{Georgios Papageorgiou\\
Department of Economics, Mathematics and Statistics\\
Birkbeck, University of London, UK\\
g.papageorgiou@bbk.ac.uk}
\date{}
\begin{document}
\maketitle
	
\begin{center}
\emph{Abstract}
\end{center}

The R package \CRANpkg{BNSP} provides a unified framework for semiparametric location-scale regression and stochastic search variable selection. 
The statistical methodology that the package is built upon utilizes basis function expansions to represent semiparametric covariate 
effects in the mean and variance functions, and spike-slab priors to perform selection and regularization of the estimated effects. 
In addition to the main function that performs posterior sampling, the package includes functions for assessing convergence of the sampler, summarizing model fits, visualizing covariate effects and obtaining predictions for new responses or their means given feature/covariate vectors.
\\


\section{Introduction}

There are many approaches to non- and semi-parametric modeling. From a Bayesian perspective, \citet{muller2013}
provide a review that covers methods for density estimation, modeling of random effects distributions in mixed
effects models, clustering and modeling of unknown functions in regression models. 

Our interest is on Bayesian methods for modeling unknown functions in regression models. 
In particular, we are interested in modeling both the mean and variance functions non-parametrically, as general 
functions of the covariates. There are multiple reasons why allowing the variance function to be a general function 
of the covariates may be important \citep{Chan06}. Firstly, it can result in more realistic prediction intervals 
than those obtained by assuming constant error variance, or as \citet{muller2013} put it, it can result in more 
honest representation of uncertainties. Secondly, it allows the practitioner to examine and understand which covariates drive the variance. 
Thirdly, it results in more efficient estimation of the mean function, and lastly, 
it produces more accurate standard errors of unknown parameters. 

In the R \citep{R} package \pkg{BNSP} \citep{bnsp} we implemented Bayesian regression 
models with Gaussian errors and with mean and log-variance functions that can be modeled as general
functions of the covariates. Covariate effects may enter the mean and log-variance functions 
parametrically or non-parametrically, with the nonparametric effects represented as linear combinations of
basis functions. The strategy that we follow in representing unknown functions is to utilize a large number 
of basis functions. This allows for flexible estimation and for capturing true effects that are locally adaptive. 
Potential problems associated with large numbers of basis functions, such as over-fitting, are avoided in our 
implementation, and efficient estimation is achieved, by utilizing spike-slab priors for variable selection.
A review of variable selection methods is provided by \citet{ohara2009}. 

The methods described here belong to the general class of models known as generalized additive models for location,
scale and shape (GAMLSS) \citep{RS05, SR07} or the Bayesian analogue termed as BAMLSS \citep{Umlauf17} and implemented 
in package \CRANpkg{bamlss} \citep{bamlss}. However, due to the nature of the spike-and-slab priors that we have implemented, 
in addition to flexible modeling of the mean and variance functions, the methods described here can also be utilized for
selecting promising subsets of predictor variables in multiple regression models. The implemented methods fall in the general 
class of methods known as stochastic search variable selection (SSVS). SSVS has received considerable attention in the Bayesian 
literature and its applications range from investigating factors that affect individual's happiness \citep{george93}, to 
constructing financial indexes \citep{george97} and to gene mapping \citep{ohara2009}. These methods associate each regression 
coefficient, either a main effect or the coefficient of a basis function, with a latent binary variable that indicates whether 
the corresponding covariate is needed in the model or not. Hence, the joint posterior distribution of the vector of these binary 
variables can identify the models with the higher posterior probability.  

R packages that are related to \pkg{BNSP} include \CRANpkg{spikeSlabGAM} \citep{ssg} that also utilizes 
SSVS methods \citep{Scheipl11}. A major difference between the two packages, however, is that whereas \pkg{spikeSlabGAM} utilizes 
spike-and-slab priors for function selection, \pkg{BNSP} utilizes spike-and-slab priors for variable selection. 
In addition, Bayesian GAMLSS models, also refer to as distributional regression models, can also be fit with R package \CRANpkg{brms} 
using normal priors \citep{brms}. Further, the R package \CRANpkg{gamboostLSS} \citep{gLSS} includes  
frequentist GAMLSS implementation based on boosting that can handle high-dimensional data \citep{Mayr12}. Lastly, the R 
package \CRANpkg{mgcv} \citep{mgcv} can also fit generalized additive models with Gaussian errors and integrated smoothness estimation, 
with implementations that can handle large datasets.  

In \pkg{BNSP} we have implemented functions for fitting such semi-parametric models, summarizing model fits, 
visualizing covariate effects and predicting new responses or their means. 
The main functions are \code{mvrm, mvrm2mcmc, print.mvrm, summary.mvrm, plot.mvrm} and \code{predict.mvrm}.
A quick description of these functions follows. 
The first one, \code{mvrm}, returns samples from the posterior distributions of the model parameters,
and it is based on an efficient Markov chain Monte Carlo (MCMC) algorithm in which we integrate out the
coefficients in the mean function, generate the variable selection indicators in blocks \citep{Chan06}
and choose the MCMC tuning parameters adaptively \citep{roberts_examples_2009}. In order to minimize random-access 
memory utilization, posterior samples are not kept in memory, but instead written in files in a directory
supplied by the user. The second function, \code{mvrm2mcmc}, reads-in the samples from the posterior 
of the model parameters and it creates an object of class \code{"mcmc"}. This enables users to
easily utilize functions from package \CRANpkg{coda} \citep{coda}, including functions \code{plot} and \code{summary}  
for assessing convergence and for summarizing posterior distributions. Further, functions \code{print.mvrm} and 
\code{summary.mvrm} provide summaries of model fits, including models and priors specified, marginal posterior 
probabilities of term inclusion in the mean and variance models and models with the highest posterior probabilities. 
Function \code{plot.mvrm} creates plots of parametric and nonparametric terms that appear in the mean 
and variance models. The function can create two-dimensional plots by calling functions from 
package \CRANpkg{ggplot2} \citep{ggplot2}. It can also create static or interactive three-dimensional plots
by calling functions from packages \CRANpkg{plot3D} \citep{plot3d} and \CRANpkg{threejs} \citep{3js}.
Lastly, function \code{predict.mvrm} provides predictions either for new responses or their means 
given feature/covariate vectors.

We next provide a detailed model description followed by illustrations on the usage of the package
and the options it provides. Technical details on the implementation of the MCMC algorithm 
are provided in the Appendix. The paper concludes with a brief summary.  

\section{Mean-variance nonparametric regression models}

Let $\uy=(y_1,\dots,y_n)^{\top}$ denote the vector of responses and let $\uX = [\ux_1,\dots,\ux_n]^{\top}$ and 
$\uZ = [\uz_1,\dots,\uz_n]^{\top}$ denote design matrices. The models that we consider express the vector of responses utilizing 
\begin{equation}
\uY = \beta_0 \uone_n + \uX \ubeta_1 + \uepsilon, \nonumber  
\end{equation}
where $\uone_n$ is the usual n-dimensional vector of ones, $\beta_0$ is an intercept term, $\ubeta_1$ is a vector of 
regression coefficients and $\uepsilon = (\epsilon_1,\dots,\epsilon_n)^{\top}$ is an n-dimensional vector of independent random
errors. Each $\epsilon_i, i=1,\dots,n,$ is assumed to have a normal distribution, $\epsilon_i \sim N(0,\sigma^2_i)$, with variances 
that are modeled in terms of covariates. 
Let $\usigma^2=(\sigma_1^2,\dots,\sigma_n^2)^{\top}$. We model the vector of variances utilizing
\begin{equation}\label{mv1}
\log(\usigma^2) = \alpha_0 \uone_n + \uZ \ualpha_1, \nonumber
\end{equation}
where $\alpha_0$ is an intercept term and $\ualpha_1$ is a vector of regression coefficients. 
Equivalently, the model for the variances can be expressed as
\begin{equation}\label{mv2}
\sigma^2_i = \sigma^2 \exp(\uz_i^{\top} \ualpha_1), i=1,\dots,n, 
\end{equation}
where $\sigma^2=\exp(\alpha_0)$. 

Let $\uD(\ualpha)$ denote an n-dimensional, diagonal matrix with elements $\exp(\uz_i^{\top} \ualpha_1/2), i=1,\dots,n$.  
Then, the model that we consider may be expressed more economically as
\begin{eqnarray}
&&\uY = \uX^* \ubeta + \uepsilon, \nonumber\\ 
&&\uepsilon \sim N(\uzero, \sigma^2 \uD^2(\ualpha)), \label{linear}
\end{eqnarray}
where $\ubeta = (\beta_0,\ubeta_1^{\top})^{\top}$ and $\uX^* = [\uone_n, \uX]$.

In the next subsections we describe how, within model (\ref{linear}), both parametric and nonparametric 
effects of explanatory variables on the mean and variance functions can be captured utilizing 
regression splines and variable selection methods. We begin by considering the special case
where there is a single covariate entering the mean model and a single covariate entering the variance 
model.

\subsection{Locally adaptive models with a single covariate}

Suppose that the observed dataset consists of triplets $(y_i,u_i,w_i), i=1,\dots,n,$ where explanatory 
variables $u$ and $w$ enter flexibly the mean and variance models, respectively. 
To model the nonparametric effects of $u$ and $w$ we consider the following formulations of the mean and variance models  
\begin{eqnarray}
&& \mu_{i} = \beta_0 + f_{\mu}(u_i) = \beta_0 + \sum_{j=1}^{q_1} \beta_{j} \phi_{1j}(u_i) = \beta_0 + \ux_i^{\top} \ubeta_1, \label{mu1}\\
&& \log(\sigma^2_{i}) = \alpha_0 + f_{\sigma}(w_i) = \alpha_0 + \sum_{j=1}^{q_2} \alpha_{j} \phi_{2j}(w_i) = \alpha_0 + \uz_i^{\top} \ualpha_1. \label{logSigma1}
\end{eqnarray}
In the preceding $\ux_i = (\phi_{11}(u_i),\dots,\phi_{1q_{1}}(u_i))^{\top}$ and 
$\uz_i = (\phi_{21}(w_i),\dots,\phi_{2q_{2}}(w_i))^{\top}$ are vectors of basis functions
and $\ubeta_1 = (\beta_1,\dots,\beta_{q_1})^{\top}$ and $\ualpha_1 = (\alpha_1,\dots,\alpha_{q_2})^{\top}$
are the corresponding coefficients. 

In package \pkg{BNSP} we implemented two sets of basis functions. Firstly, radial basis functions
\begin{eqnarray}\label{rbf}
\mathcal{B}_1 = \big\{\phi_{1}(u)=u , \phi_{2}(u)=||u-\xi_{1}||^2 \log\left(||u-\xi_{1}||^2\right), \dots,\nonumber\\ 
\phi_{q}(u)=||u-\xi_{q-1}||^2 \log\left(||u-\xi_{q-1}||^2\right)\big\},
\end{eqnarray}
where $||u||$ denotes the Euclidean norm of $u$ and $\xi_1,\dots,\xi_{q-1}$ are the knots that within package 
\pkg{BNSP} are chosen as the quantiles of the observed values of explanatory variable $u$, 
with $\xi_1=\min(u_i)$, $\xi_{q-1}=\max(u_i)$ and the remaining knots chosen as equally spaced quantiles between
$\xi_1$ and $\xi_{q-1}$. 

Secondly, we implemented thin plate splines 
\begin{equation}
\mathcal{B}_2 = \left\{\phi_{1}(u)=u , \phi_{2}(u)=(u-\xi_{1})_{+}, \dots, \phi_{q}(u)=(u-\xi_{q})_{+}\right\},\nonumber
\end{equation}
where $(a)_+ = \max(a,0)$ and the knots $\xi_1,\dots,\xi_{q-1}$ are determined as above. 
 
In addition, \pkg{BNSP} supports the smooth constructors from package \pkg{mgcv} e.g. the low-rank thin plate splines,
cubic regression splines, P-splines, their cyclic versions and others. Examples on how these smooth terms are used within \pkg{BNSP} 
are provided later in this paper.   

Locally adaptive models for the mean and variance functions are obtained utilizing the methodology 
developed by \citet{Chan06}. Considering the mean function, local adaptivity is achieved by 
utilizing a large number of basis functions $q_1$. Over-fitting, and problems associated with it, is 
avoided by allowing positive prior probability that the regression coefficients are exactly zero.
The latter is achieved by defining binary variables $\gamma_{j}, j=1,\dots,q_1,$ that 
take value $\gamma_{j} = 1$ if $\beta_{j} \neq 0$ and $\gamma_{j} = 0$ if $\beta_j = 0$. 
Hence, vector $\ugamma = (\gamma_{1}, \dots, \gamma_{q_1})^{\top}$ determines which terms enter the mean model. 
The vector of indicators $\udelta = (\delta_{1}, \dots, \delta_{q_2})^{\top}$ for the variance function is defined analogously. 

Given vectors $\ugamma$ and $\udelta$, the heteroscedastic, semiparametric model (\ref{linear}) can be written as 
\begin{eqnarray}\label{modcor}
&&\uY = \uX^*_{\gamma} \ubeta_{\gamma} + \uepsilon,\nonumber\\
&&\uepsilon \sim N(\uzero, \sigma^2 \uD^2(\ualpha_{\delta})),\nonumber
\end{eqnarray}
where $\ubeta_{\gamma}$ consisting of all non-zero elements of $\ubeta_1$
and $\uX^*_{\gamma}$ consists of the corresponding columns of $\uX^*$. 
Subvector $\ualpha_{\delta}$ is defined analogously. 

We note that, as was suggested by \citet{Chan06}, we work with mean corrected columns in the design matrices
$\uX$ and $\uZ$, both in this paper and in the \pkg{BNSP} implementation. We remove the mean from all
columns in the design matrices except those that correspond to categorical variables. 

\subsection{Prior specification for models with a single covariate}\label{prior}

Let $\;\utX = \uD(\ualpha)^{-1} \uX^*$. 
The prior for $\ubeta_{\gamma}$ is specified as \citep{AZ}
\begin{eqnarray}
\ubeta_{\gamma} | c_{\beta}, \sigma^2, \ugamma, \ualpha, \udelta \sim N(\uzero,c_{\beta} \sigma^2 (\utX_{\gamma}^{\top} \utX_{\gamma} )^{-1}).\nonumber
\end{eqnarray}

Further, the prior for $\ualpha_{\delta}$ is specified as
\begin{eqnarray}
\ualpha_{\delta} | c_{\alpha}, \udelta \sim N(\uzero,c_{\alpha} \uI).\nonumber
\end{eqnarray}

Independent priors are specified for the indicators variables $\gamma_j$ as 
$P(\gamma_{j} = 1 | \pi_{\mu}) = \pi_{\mu}, j=1,\dots,q_1$, from which the 
joint prior is obtained as
\begin{eqnarray}
P(\ugamma|\pi_{\mu}) = \pi_{\mu}^{N(\gamma)} (1-\pi_{\mu})^{q_1-N(\gamma)},\nonumber
\end{eqnarray}
where $N(\gamma) = \sum_{j=1}^{q_1} \gamma_j$.

Similarly, for the indicators $\delta_j$ we specify independent priors 
$P(\delta_{j} = 1 | \pi_{\sigma}) = \pi_{\sigma}, j=1,\dots,q_2$. It follows that the joint prior
is 
\begin{eqnarray}
P(\udelta|\pi_{\sigma}) = \pi_{\sigma}^{N(\delta)} (1-\pi_{\sigma})^{q_2-N(\delta)},\nonumber
\end{eqnarray}
where $N(\delta) = \sum_{j=1}^{q_2} \delta_j$.

We specify inverse Gamma priors for $c_{\beta}$ and $c_{\alpha}$ and Beta priors for 
$\pi_{\mu}$ and $\pi_{\sigma}$
\begin{eqnarray}\label{PriorParams1}
&& c_{\beta} \sim \text{IG}(a_{\beta},b_{\beta}),
c_{\alpha} \sim \text{IG}(a_{\alpha},b_{\alpha}),\nonumber\\
&& \pi_{\mu} \sim \text{Beta}(c_{\mu},d_{\mu}), \pi_{\sigma} \sim \text{Beta}(c_{\sigma},d_{\sigma}).
\end{eqnarray}
Lastly, for $\sigma^2$ we consider inverse Gamma and half-normal priors
\begin{eqnarray}\label{PriorParams2}
\sigma^2 \sim \text{IG}(a_{\sigma},b_{\sigma}) \;\text{and}\; \sigma \sim N(\sigma;0,\phi^2_{\sigma}) I[\sigma > 0].
\end{eqnarray}

\subsection{Extension to bivariate covariates}

It is straightforward to extend the methodology described earlier 
to allow fitting of flexible mean and variance surfaces. In fact, the only modification required is in
the basis functions and knots. For fitting surfaces, in package \pkg{BNSP} we implemented 
radial basis functions
\begin{eqnarray}
\mathcal{B}_3 = \Big\{\phi_{1}(\uu)=u_1,\phi_{2}(\uu)=u_2,\phi_{3}(\uu)=||\uu-\uxi_{1}||^2 \log\left(||\uu-\uxi_{1}||^2\right),\dots,
\nonumber \\
\phi_{q}(\uu)=||\uu-\uxi_{q-2}||^2 \log\left(||\uu-\uxi_{q-2}||^2\right)\Big\}.\nonumber
\end{eqnarray}

We note that the prior specification presented earlier for fitting flexible functions 
remains unchained for fitting flexible surfaces. Further, for fitting bivariate or higher order functions, 
\pkg{BNSP} also supports smooth constructors \code{s}, \code{te} and \code{ti} from \pkg{mgcv}.

\subsection{Extension to additive models}

In the presence of multiple covariates, the effects of which may be modeled parametrically or
semiparametrically, the mean model in (\ref{mu1}) is extended to the following
\begin{eqnarray}
\mu_{i} = \beta_0 + \uu_{ip}^{\top} \ubeta  + \sum_{k=1}^{K_1} f_{\mu,k}(u_{ik}), i=1,\dots,n,\nonumber
\end{eqnarray}
where, $\uu_{ip}$ includes the covariates the effects of which are modeled parametrically,  
$\ubeta$ denotes the corresponding effects, and $f_{\mu,k}(u_{ik}), k=1,\dots,K_1,$ are flexible functions
of one or more covariates expressed as
\begin{equation}
f_{\mu,k}(u_{ik}) = \sum_{j=1}^{q_{1k}} \beta_{kj} \phi_{1kj}(u_{ik}),\nonumber
\end{equation}
where $\phi_{1kj}, j=1,\dots,q_{1k}$ are the basis functions used in the $k$th component, 
$k=1,\dots,K_1$.  

Similarly, the variance model (\ref{logSigma1}), in the presence of multiple covariates,
is expressed as 
\begin{eqnarray}
\log(\sigma^2_i) = \alpha_0 + \uw_{ip}^{\top} \ualpha  + \sum_{k=1}^{K_2} f_{\sigma,k}(w_{ik}), i=1,\dots,n,\nonumber
\end{eqnarray}
where
\begin{equation}
f_{\sigma,k}(w_{ik}) = \sum_{j=1}^{q_{2k}} \alpha_{kj} \phi_{2kj}(w_{ik}).\nonumber
\end{equation}

For additive models, local adaptivity is achieved using a similar strategy as in the single covariate case.
That is, we utilize a potentially large number of knots or basis functions in the flexible components that 
appear in the mean model, $f_{\mu,k},k=1,\dots,K_1,$ and 
in the variance model, $f_{\sigma,k},k=1,\dots,K_2$. To avoid over-fitting, we allow removal of the 
unnecessary ones utilizing the usual indicator variables,
$\ugamma_k = (\gamma_{k1},\dots,\gamma_{kq_{1k}})^{\top}, k=1,\dots,K_1,$ and
$\udelta_k = (\delta_{k1},\dots,\delta_{kq_{2k}})^{\top}, k=1,\dots,K_2.$ 
Here, vectors $\ugamma_k$ and $\udelta_k$ determine which basis functions appear in $f_{\mu,k}$
and $f_{\sigma,k}$ respectively. 

The model that we implemented in package \pkg{BNSP} specifies independent priors for the indicators variables $\gamma_{kj}$ as 
$P(\gamma_{kj} = 1 | \pi_{\mu_k}) = \pi_{\mu_k}, j=1,\dots,q_{1k}$. From these, the joint prior follows 
\begin{eqnarray}
P(\ugamma_k|\pi_{\mu_k}) = \pi_{\mu_k}^{N(\gamma_k)} (1-\pi_{\mu_k})^{q_{1k}-N(\gamma_k)},\nonumber
\end{eqnarray}
where $N(\gamma_k) = \sum_{j=1}^{q_{1k}} \gamma_{kj}$.

Similarly, for the indicators $\delta_{kj}$ we specify independent priors 
$P(\delta_{kj} = 1 | \pi_{\sigma_k}) = \pi_{\sigma_k}, j=1,\dots,q_{2k}$. It follows that the joint prior
is 
\begin{eqnarray}
P(\udelta_k|\pi_{\sigma_k}) = \pi_{\sigma_k}^{N(\delta_k)} (1-\pi_{\sigma_k})^{q_{2k}-N(\delta_k)},\nonumber
\end{eqnarray}
where $N(\delta_k) = \sum_{j=1}^{q_{2k}} \delta_{kj}$.

We specify the following independent priors for the inclusion probabilities.  
\begin{equation}
\pi_{\mu_k} \sim \text{Beta}(c_{\mu_k},d_{\mu_k}), k=1,\dots,K_1 \quad
\pi_{\sigma_k} \sim \text{Beta}(c_{\sigma_k},d_{\sigma_k}), k=1,\dots,K_2. \label{gam.priors}
\end{equation}

The rest of the priors are the same as those specified for the single covariate models. 

\section{Usage}\label{usage}

In this section we provide results on simulation studies and real data analyses. The 
purpose is twofold: firstly we point out that the package works well and provides the
expected results (in simulation studies) and secondly we illustrate the options that the
users of \pkg{BNSP} have.

\subsection{Simulation studies}
 
Here we present results from three simulations studies, involving one, two, and multiple
covariates. For the majority of these simulation studies, we utilize the 
same data-generating mechanisms as those presented by \citet{Chan06}. 
 
\subsubsection{Single covariate case} 
 
We consider two mechanisms that involve a single covariate that appears in both 
the mean and variance model. Denoting the covariate by $u$, the data-generating mechanisms are the normal model
$Y \sim N\{\mu(u),\sigma^2(u)\}$ with the following mean and standard deviation functions 
\begin{enumerate}
\item $\mu(u) = 2u, \sigma(u)=0.1+u$,
\item $\mu(u) = \left\{N(u,\mu=0.2,\sigma^2=0.004)+N(u,\mu=0.6,\sigma^2=0.1)\right\}/4,$ \\
      $\sigma(u)=\left\{N(u,\mu=0.2,\sigma^2=0.004)+N(u,\mu=0.6,\sigma^2=0.1)\right\}/6$.
\end{enumerate}
We generate a single dataset of size $n=500$ from each mechanism, 
where variable $u$ is obtained from the uniform distribution, $u \sim \text{Unif}(0,1)$.
For instance, for obtaining a dataset from the first mechanism we use 
\begin{verbatim}
> mu <- function(u){2 * u}
> stdev <- function(u){0.1 + u}
> set.seed(1)
> n <- 500
> u <- sort(runif(n))
> y <- rnorm(n,mu(u),stdev(u))
> data <- data.frame(y,u)
\end{verbatim}
Above we specified the seed value to be one, and we do so in what follows, so that our results are replicable. 
 
To the generated dataset we fit a special case of the model that we presented, where the mean and variance 
functions in (\ref{mu1}) and (\ref{logSigma1}) are specified as
\begin{eqnarray}
\mu = \beta_0 + f_{\mu}(u) = \beta_0 + \sum_{j=1}^{q_1} \beta_{j} \phi_{1j}(u) \quad\text{and}\quad
\log(\sigma^2) = \alpha_0 + f_{\sigma}(u) = \alpha_0 + \sum_{j=1}^{q_2} \alpha_{j} \phi_{2j}(u), \label{modelsim1}
\end{eqnarray}
with $\phi$ denoting the radial basis functions presented in (\ref{rbf}). Further, we choose $q_1=q_2=21$ basis functions 
or, equivalently, $20$ knots. Hence, we have $\phi_{1j}(u)=\phi_{2j}(u), j=1,\dots,21,$ which results in identical 
design matrices for the mean and variance models. In R, the two models are specified using  
\begin{verbatim}
> model <- y ~ sm(u, k = 20, bs = "rd") | sm(u, k = 20, bs = "rd")
\end{verbatim}
The above formula \citep{ZC10} specifies the response, mean and variance models.
Smooth terms are specified utilizing function \code{sm}, that takes as input the covariate $u$, 
the selected number of knots and the selected type of basis functions.

Next we specify the hyper-parameter values for the priors in (\ref{PriorParams1}) and (\ref{PriorParams2}). 
The default prior for $c_{\beta}$ is inverse Gamma with $a_{\beta}=0.5,b_{\beta}=n/2$ \citep{liang_mixtures_2008}. 
For parameter $c_{\alpha}$ the default prior is a non-informative but proper inverse Gamma 
with $a_{\alpha}=b_{\alpha}=1.1$. Concerning $\pi_{\mu}$ and $\pi_{\sigma}$, the default priors are 
uniform, obtained by setting $c_{\mu}=d_{\mu}=1$ and $c_{\sigma}=d_{\sigma}=1$. Lastly, the 
default prior for the error standard deviation is the half-normal with variance $\phi^2_{\sigma}=2$, 
$|\sigma|\sim N(0,2)$.  

We choose to run the MCMC sampler for $10,000$ iterations and discard the first $5,000$ as burn-in.
Of the remaining $5,000$ samples we retain 1 every 2 samples, resulting in $2,500$ posterior samples.
Further, as mentioned above, we set the seed of the MCMC sampler equal to one. 
Obtaining posterior samples is achieved by a function call of the form
\begin{verbatim}
> m1 <- mvrm(formula = model, data = data, sweeps = 10000, burn = 5000, 
+       thin = 2, seed = 1, StorageDir = DIR,
+       c.betaPrior = "IG(0.5,0.5*n)", c.alphaPrior = "IG(1.1,1.1)",
+       pi.muPrior = "Beta(1,1)", pi.sigmaPrior = "Beta(1,1)", sigmaPrior = "HN(2)")
\end{verbatim}
Samples from the 
posteriors of the model parameters $\{\ubeta,\ugamma,\ualpha,\udelta,c_{\beta},c_{\alpha},\sigma^2\}$ are written
in seven separate files which are stored in the directory specified by argument \code{StorageDir}. If a storage directory
is not specified, then function \code{mvrm} returns an error message, as without these files there will be 
no output to process. Furthermore, the last two lines of the above function call show the specified priors, which are 
$c_{\beta} \sim \text{IG}(0.5,n/2)$, $c_{\alpha} \sim \text{IG}(1.1,1.1),$ 
$\pi_{\mu} \sim \text{Beta}(1,1)$, $\pi_{\sigma} \sim \text{Beta}(1,1)$ and $|\sigma|\sim N(0,2)$, respectively.  
As we mentioned above, these priors are the default ones, and hence the same function call can be achieved 
without specifying the last two lines. Here we display the priors in order to describe how users
can specify their own priors. For parameters $c_{\beta}$ and $c_{\alpha}$ only inverse Gamma priors are available,
with parameters that can be specified by the user in the intuitive way. 
For instance, the prior $c_{\beta} \sim \text{IG}(1.01,1.01)$ can be specified in function
\code{mvrm} by using \code{c.betaPrior = "IG(1.01,1.01)"}. The second parameter of the prior for $c_{\beta}$ can be 
a function of the sample size $n$ (but only symbol $n$ would work here), so for instance \code{c.betaPrior = "IG(1,0.4*n)"}
is another acceptable specification. Further, Beta priors are available for 
parameters $\pi_{\mu}$ and $\pi_{\sigma}$ with parameters that can be specified by the user again in the intuitive
way. Lastly, two priors are available for the error variance. These are the default half-normal and the inverse Gamma. 
For instance, \code{sigmaPrior = "HN(5)"} defines $|\sigma|\sim N(0,5)$ as the prior while \code{sigmaPrior = "IG(1.1,1.1)"}  
defines $\sigma^2 \sim \text{IG}(1.1,1.1)$ as the prior.

Function \code{mvrm2mcmc} reads in posterior samples from the files that the call to function \code{mvrm}
generated and creates an object of class \code{"mcmc"}. Hence, for summarizing posterior distributions and for assessing convergence, 
functions \code{summary} and \code{plot} from package \pkg{coda} can be used. As an example, here
we read in the samples from the posterior of $\ubeta$
\begin{verbatim}
> beta <- mvrm2mcmc(m1, "beta")
\end{verbatim}
and summarize the posterior using \code{summary}.
For the sake of economizing space, only the part of the output that describes the posteriors of 
$\beta_0, \beta_1$, and $\beta_2$ is shown below 
\begin{verbatim}
> summary(beta)

Iterations = 5001:9999
Thinning interval = 2 
Number of chains = 1 
Sample size per chain = 2500 

1. Empirical mean and standard deviation for each variable,
   plus standard error of the mean:

                  Mean       SD  Naive SE Time-series SE
(Intercept)  9.534e-01 0.004399 8.799e-05      0.0002534
u            1.864e+00 0.042045 8.409e-04      0.0010356
sm(u).1      3.842e-04 0.016421 3.284e-04      0.0003284


2. Quantiles for each variable:

              2.5%    25%    50%    75%  97.5%
(Intercept) 0.946 0.9513 0.9533 0.9554 0.960
u           1.833 1.8565 1.8614 1.8682 1.923
sm(u).1     0.000 0.0000 0.0000 0.0000 0.000
\end{verbatim}

Further, we may obtain a plot using 
\begin{verbatim}
> plot(beta)
\end{verbatim}

\begin{figure}
\begin{center}
\includegraphics[height=0.3\textheight]{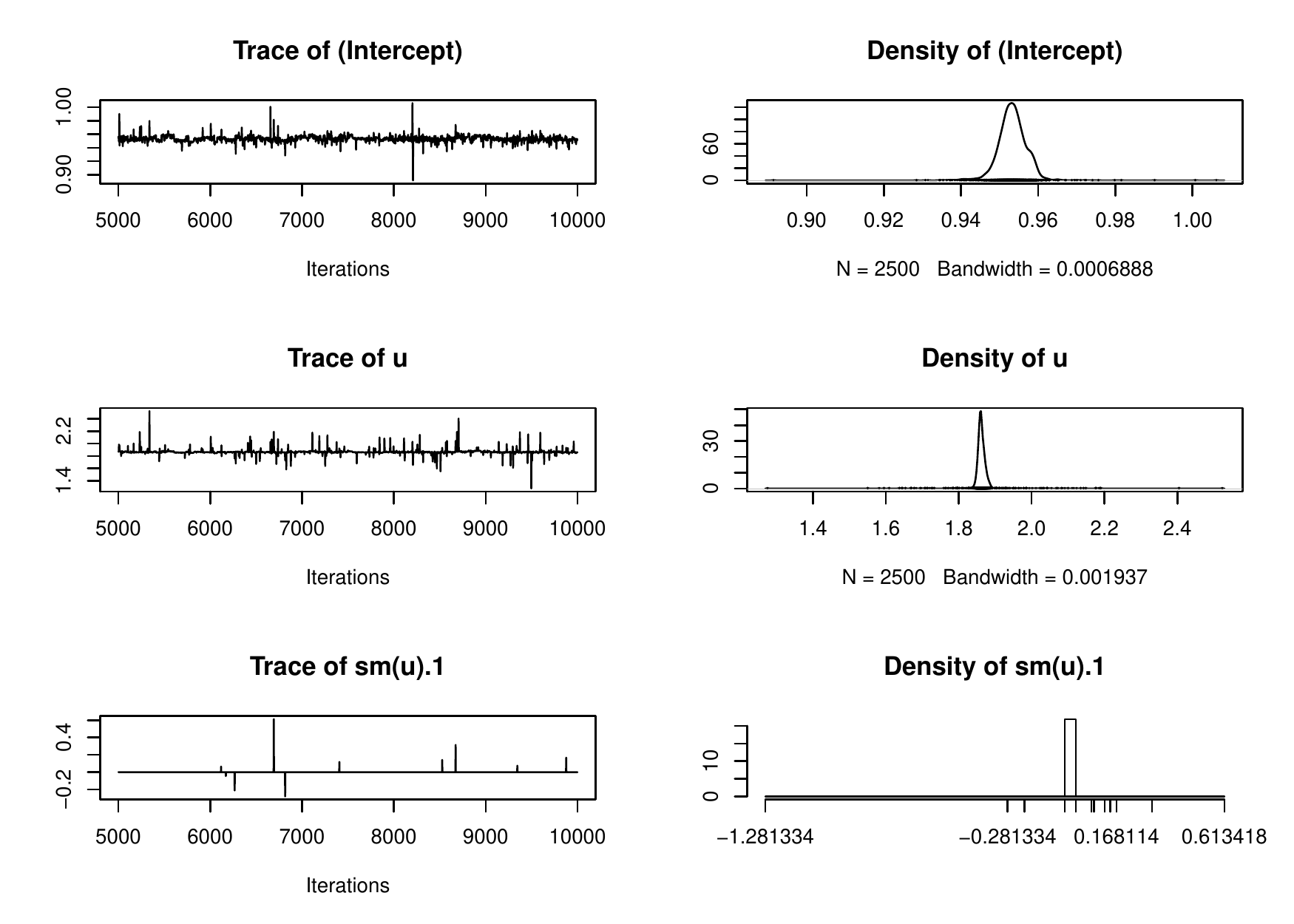}
\end{center}
\caption{Trace and density plots for the regression coefficients $\beta_0, \beta_1$ and $\beta_2$ of the first 
simulated example. Parameters $\beta_1$ and $\beta_2$ are the coefficients of the first two basis functions, 
denoted by `u' and `sm(u).1'. Plots for coefficients 
$\beta_3,\dots,\beta_{21}$ are omitted as they follow a very similar pattern to that seen for $\beta_2$ 
i.e. most of the time they take value zero but with random spikes away from zero.}\label{betasim1}
\end{figure}

Figure~\ref{betasim1} shows the first three of the plots created by function \code{plot}. These are the plots
of the samples from the posteriors of coefficients $\beta_0, \beta_1$ and $\beta_2$.
As we can see from both the summary and Figure~\ref{betasim1}, only the first two coefficients have posteriors 
that are not centered around zero. 

Returning to function \code{mvrm2mcmc}, it requires two inputs. These are an object of class \code{"mvrm"} and the name of the file
to be read in R. For the parameters in the current model $\{\ubeta,\ugamma,\ualpha,\udelta,c_{\beta},c_{\alpha},\sigma^2\}$
the corresponding file names are `beta', `gamma', `alpha', `delta', `cbeta', `calpha' and `sigma2' respectively. 

Summaries of `mvrm' fits may be obtained utilizing functions \code{print.mvrm} and \code{summary.mvrm}. 
Function \code{print} takes as input an object of class \code{"mvrm"}. It returns basic information at 
the model fit as shown below
\begin{verbatim}
> print(m1)

Call:
mvrm(formula = model, data = data, sweeps = 10000, burn = 5000, 
    thin = 2, seed = 1, StorageDir = DIR, c.betaPrior = "IG(0.5,0.5*n)", 
    c.alphaPrior = "IG(1.1,1.1)", pi.muPrior = "Beta(1,1)", 
    pi.sigmaPrior = "Beta(1,1)", sigmaPrior = "HN(2)")

2500 posterior samples

Mean model - marginal inclusion probabilities
       u  sm(u).1  sm(u).2  sm(u).3  sm(u).4  sm(u).5  sm(u).6  sm(u).7 
  1.0000   0.0040   0.0036   0.0032   0.0084   0.0036   0.0044   0.0028 
 sm(u).8  sm(u).9 sm(u).10 sm(u).11 sm(u).12 sm(u).13 sm(u).14 sm(u).15 
  0.0060   0.0020   0.0060   0.0036   0.0056   0.0056   0.0036   0.0052 
sm(u).16 sm(u).17 sm(u).18 sm(u).19 sm(u).20 
  0.0060   0.0044   0.0056   0.0044   0.0052 

Variance model - marginal inclusion probabilities
       u  sm(u).1  sm(u).2  sm(u).3  sm(u).4  sm(u).5  sm(u).6  sm(u).7 
  1.0000   0.6072   0.5164   0.5808   0.5488   0.6760   0.5320   0.6336 
 sm(u).8  sm(u).9 sm(u).10 sm(u).11 sm(u).12 sm(u).13 sm(u).14 sm(u).15 
  0.6936   0.6708   0.5996   0.4816   0.4912   0.3728   0.6268   0.5688 
sm(u).16 sm(u).17 sm(u).18 sm(u).19 sm(u).20 
  0.5872   0.6528   0.4428   0.6900   0.5356
\end{verbatim}

The function returns a matched call, the number of posterior samples obtained
and marginal inclusion probabilities of the terms in the mean and variance models. 

\begin{figure}
\begin{center}
\begin{tabular}{cc}
\includegraphics[width=0.4\textwidth,height=0.17\textheight]{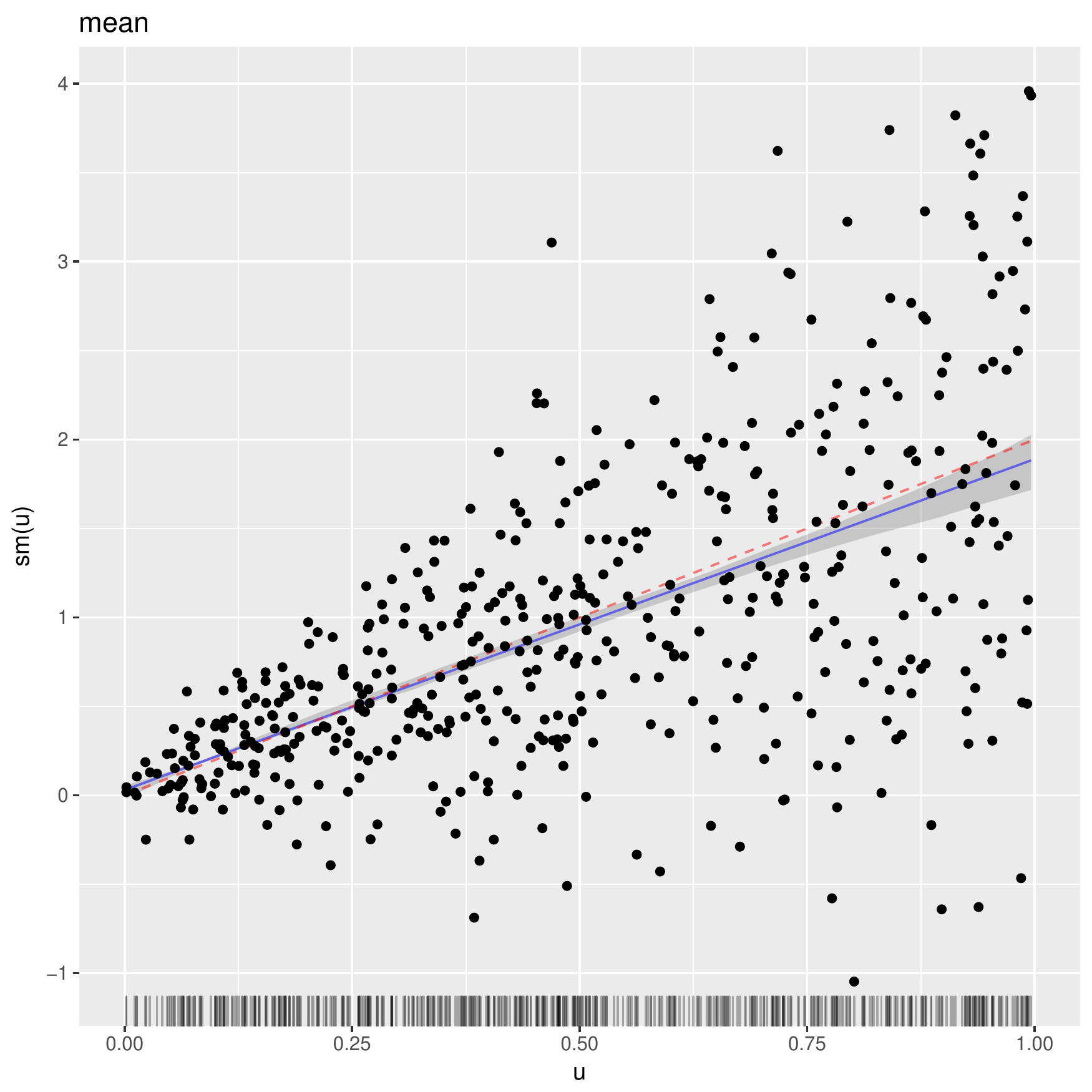} &  
\includegraphics[width=0.4\textwidth,height=0.17\textheight]{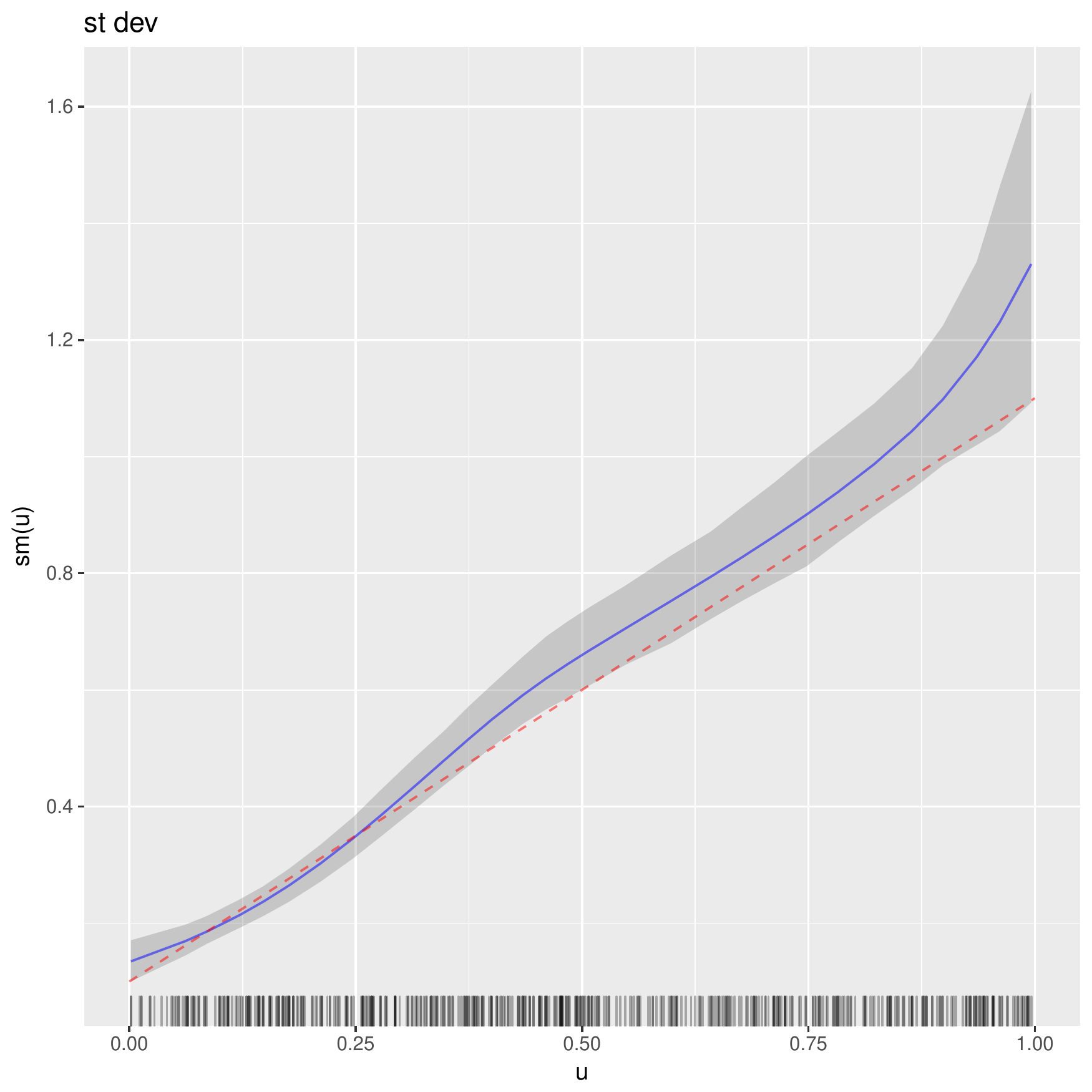} \\
(a) & (b) \\
\includegraphics[width=0.4\textwidth,height=0.17\textheight]{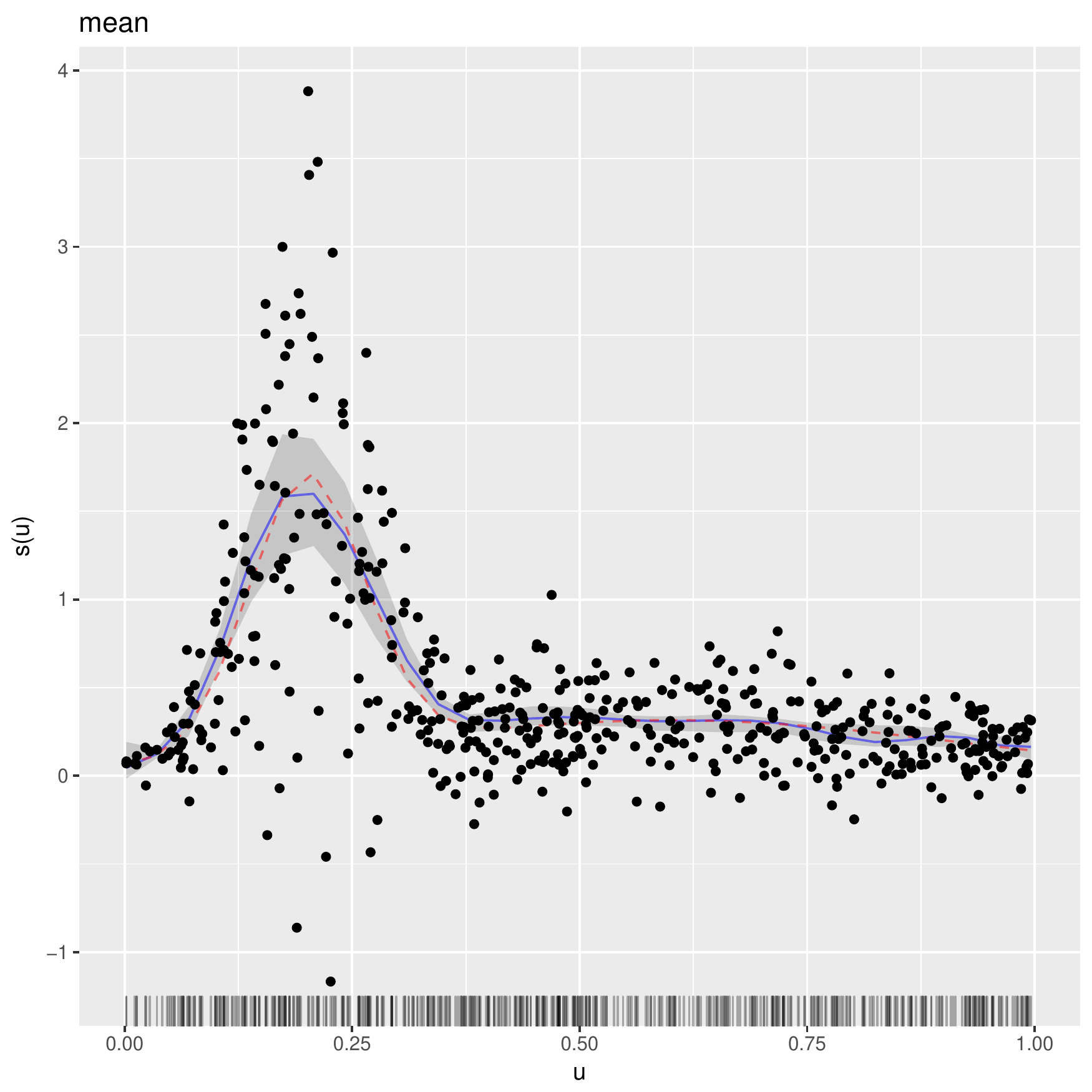} &  
\includegraphics[width=0.4\textwidth,height=0.17\textheight]{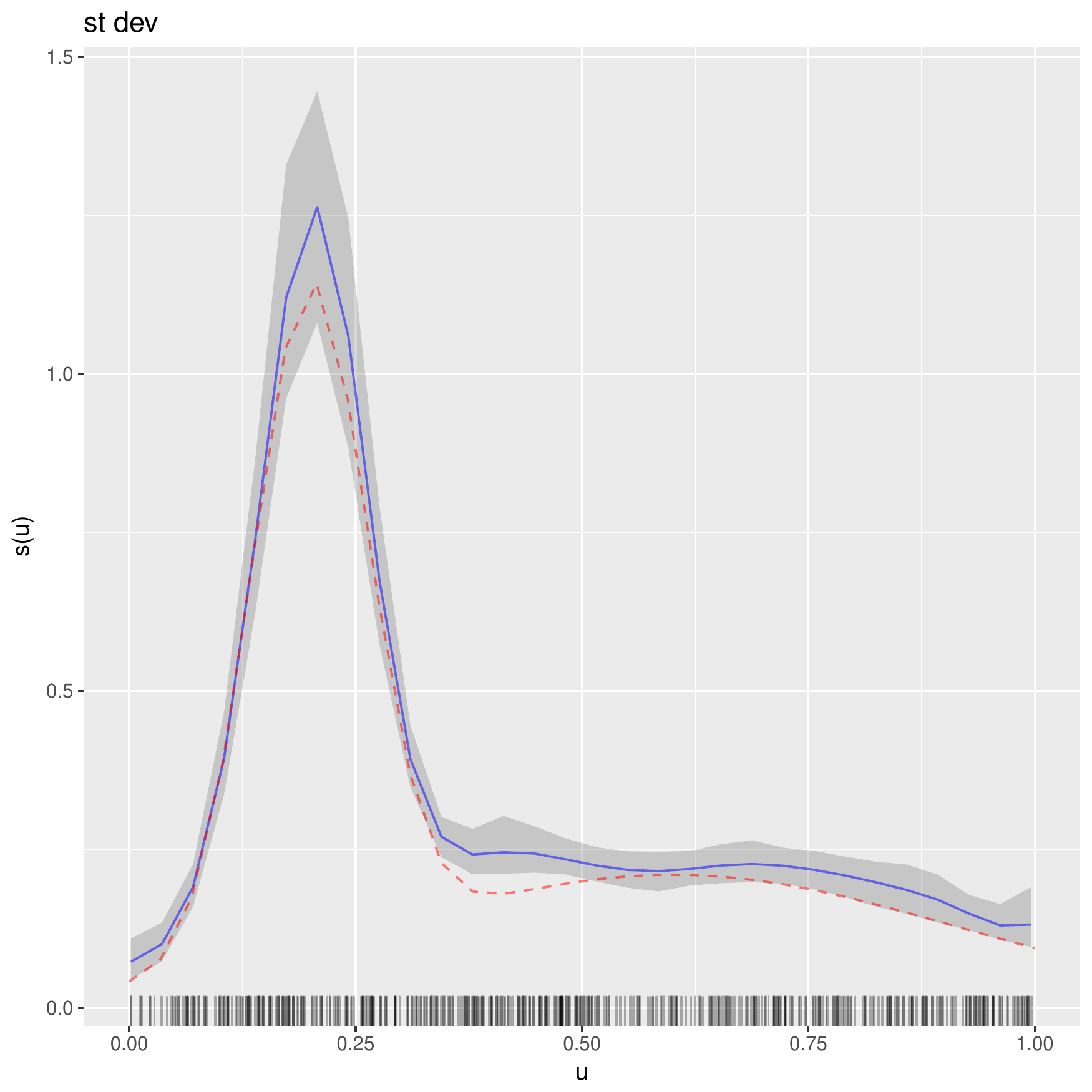} \\
(c) & (d) \\
\end{tabular}
\end{center}
\caption{Results from the single covariate simulated examples. The column on the left-hand side
displays the generated data points and posterior means of the estimated effect along with $99\%$ CIs. 
The column on the right-hand side displays the posterior mean of the estimated standard deviation function along with $90\%$ CIs. 
In all panels, the truth is represented by dashed (red color) lines, the posterior means by solid 
(blue color) lines, and the CIs by gray color.}\label{sim1mv}
\end{figure}

Whereas the output of function \code{print} focuses on marginal inclusion probabilities,
the output of function \code{summary} focuses on the most frequently visited models.
It takes as input an object of class `mvrm' and the number of (most frequently visited) models to be displayed,
which by default is set to \code{nModels = 5}. Here to economize space we set \code{nModels = 2}.
The information returned by functions \code{summary} is shown below
\begin{verbatim}
> summary(m1, nModels = 2)

Specified model for the mean and variance:
y ~ sm(u, k = 20, bs = "rd") | sm(u, k = 20, bs = "rd")

Specified priors:
[1] c.beta = IG(0.5,0.5*n) c.alpha = IG(1.1,1.1)  pi.mu = Beta(1,1)     
[4] pi.sigma = Beta(1,1)   sigma = HN(2)         

Total posterior samples: 2500 ; burn-in: 5000 ; thinning: 2 

Files stored in /home/papgeo/1/ 
                                 
Null deviance:           1299.292
Mean posterior deviance:  -88.691

Joint mean/variance model posterior probabilities:
  mean.u mean.sm.u..1 mean.sm.u..2 mean.sm.u..3 mean.sm.u..4 mean.sm.u..5
1      1            0            0            0            0            0
2      1            0            0            0            0            0
  mean.sm.u..6 mean.sm.u..7 mean.sm.u..8 mean.sm.u..9 mean.sm.u..10
1            0            0            0            0             0
2            0            0            0            0             0
  mean.sm.u..11 mean.sm.u..12 mean.sm.u..13 mean.sm.u..14 mean.sm.u..15
1             0             0             0             0             0
2             0             0             0             0             0
  mean.sm.u..16 mean.sm.u..17 mean.sm.u..18 mean.sm.u..19 mean.sm.u..20 var.u
1             0             0             0             0             0     1
2             0             0             0             0             0     1
  var.sm.u..1 var.sm.u..2 var.sm.u..3 var.sm.u..4 var.sm.u..5 var.sm.u..6
1           1           1           1           1           1           1
2           1           0           1           1           1           1
  var.sm.u..7 var.sm.u..8 var.sm.u..9 var.sm.u..10 var.sm.u..11 var.sm.u..12
1           1           1           1            0            1            0
2           1           1           1            1            1            1
  var.sm.u..13 var.sm.u..14 var.sm.u..15 var.sm.u..16 var.sm.u..17 var.sm.u..18
1            1            1            1            1            1            1
2            0            1            1            1            1            0
  var.sm.u..19 var.sm.u..20 freq prob cumulative
1            1            1  141 5.64       5.64
2            1            0  120 4.80      10.44
Displaying 2 models of the 916 visited
2 models account for 10.44% of the posterior mass
\end{verbatim}
Firstly, the function provides the specified mean and variance models and the specified priors. 
This is followed by information about the MCMC chain and the directory where files have been  
stored. In addition, the function provides the null and the mean posterior deviance. 
Finally, the function provides the specification of the joint mean/variance models that were visited most often 
during MCMC sampling. This specification is in terms of a vector of indicators, each consisting 
of zeros and ones that show which terms are in the mean and variance model. To make
clear which terms pertain to the mean and which to the variance function, we have preceded
the names of the model terms by `\code{mean.}' or a `\code{var.}'. In the above output
we see that the most visited model specifies a linear mean model (only the linear term in included in the model)
while the variance model includes twelve terms. See also Figure~\ref{sim1mv}. 

We next describe function \code{plot.mvrm} which creates
plots of terms in the mean and variance functions. Two calls to function
\code{plot} can be seen in the code below. Argument \code{x} expects an object of class `mvrm', as created by a call to function
\code{mvrm}. Argument \code{model} may take on one of three possible values: `mean', `stdev' or `both', specifying
the model to be visualized. Further, argument \code{term} determines the term to be plotted. In the current example there is 
only one term in each of the two models which leaves us with only one choice, \code{term = "sm(u)"}. Equivalently,
\code{term} may be specified as an integer, \code{term = 1}. If term is left unspecified, then by default the first term in the 
model is plotted. For creating two-dimensional plots, as in the current example, function \code{plot}  
utilizes package \pkg{ggplot2}. Users of \pkg{BNSP} may add their own options to plots
via argument \code{plotOptions}. The code below serves as an example.
\begin{verbatim}
> x1 <- seq(0, 1, length.out = 30)
> plotOptionsM <- list(geom_line(aes_string(x = x1, y = mu(x1)), col = 2, alpha = 0.5, 
+                      lty = 2), geom_point(data = data, aes(x = u, y = y)))
> plot(x = m1, model = "mean", term = "sm(u)", plotOptions = plotOptionsM, 
+      intercept = TRUE, quantiles = c(0.005, 0.995), grid = 30)
> plotOptionsV = list(geom_line(aes_string(x = x1, y = stdev(x1)), col = 2, 
+                     alpha = 0.5, lty = 2))
> plot(x = m1, model = "stdev", term = "sm(u)", plotOptions = plotOptionsV, 
+      intercept = TRUE, quantiles = c(0.05, 0.95), grid = 30) 
\end{verbatim}

The resulting plots can be seen in Figure~\ref{sim1mv}, panels (a) and (b). Panel (a) displays the simulated dataset, showing the expected 
increase in both the mean and variance with increasing values of the covariate $u$. Further, we see the posterior 
mean of $\mu(u) = \beta_0 + f_{\mu}(u) = \beta_0 + \sum_{j=1}^{21} \beta_{j} \phi_{1j}(u)$ evaluated over a grid of $30$ values of
$u$, as specified by the (default) \code{grid = 30} option in function \code{plot}.
For each sample $\ubeta^{(s)}, s=1,\dots,S,$ from the posterior of $\ubeta$, and for each value of $u$ over the grid of $30$ values,
$u_j, j=1,\dots,30,$ function \code{plot} computes $\mu(u_j)^{(s)} = \beta_0^{(s)} + \sum_{j=1}^{21} \beta_{j}^{(s)} \phi_{1j}(u_j)$.
The default option \code{intercept = TRUE} specifies that the intercept $\beta_0$ is included in the computation, but it may
be removed by setting \code{intercept = FALSE}. 
The posterior means are computed by the usual $\bar{\mu}(u_j) = S^{-1}\sum_{s} \mu(u_j)^{(s)}$ and are plotted with
solid (blue-color) line. By default, the function displays $80\%$ point-wise credible intervals (CI). In Figure~\ref{sim1mv} 
panel (a) we have plotted $99\%$ CIs, as specified by option \code{quantiles = c(0.005, 0.995)}. This option specifies 
that for each value $u_j, j=1,\dots,30,$ on the grid, $99\%$ CIs for $\mu(u_j)$ are computed by the $0.5\%$ and $99.5\%$ quantiles 
of the samples $\mu(u_j)^{(s)}, s=1,\dots,S$.      
Plots without credible intervals may be obtained by setting \code{quantiles = NULL}.

Figure~\ref{sim1mv}, panel (b) displays the posterior mean of the standard deviation function, given by 
$\sigma(u) = \sigma \exp\{\sum_{j=1}^{q_2} \alpha_{j} \phi_{2j}(u)/2\}$. The details are the same as for the plot 
of the mean function, so here we briefly mention a difference: option \code{intercept = TRUE} specifies that $\sigma$ 
is included in the calculation. It may be removed by setting \code{intercept = FALSE}, which will result in plots of 
$\sigma(u)^{*} = \exp\{\sum_{j=1}^{q_2} \alpha_{j} \phi_{2j}(u)/2\}$. 

We use the second simulated dataset to show how the \code{s} constructor from package \pkg{mgcv} may be used. 
In our example, we used \code{s} to specify the model as follows
\begin{verbatim}
> model <- y ~ s(u, k = 15, bs = "ps", absorb.cons=TRUE) | 
+              s(u, k = 15, bs = "ps", absorb.cons=TRUE)
\end{verbatim}
Function \code{BNSP::s} calls in turn \code{mgcv::s} and \code{mgcv::smoothCon}. All options of the last two functions 
may be passed to \code{BNSP::s}. In the example above we used options \code{k}, \code{bs} and \code{absorb.cons}.        

The remaining R code for the second simulated example is precisely the same as the one for the first example, and hence omitted. 
Results are shown in Figure~\ref{sim1mv}, panels (c) and (d).  

We conclude the current section by describing the function \code{predict.mvrm}. The function provides predictions 
and posterior credible or prediction intervals for given feature vectors. The two types of intervals
differ in the associated level of uncertainty: prediction intervals attempt to capture a future response and are usually much
wider than credible intervals that attempt to capture a mean response. 

The following code shows how credible and prediction intervals can be obtained for a sequence of covariate
values stored in \code{x1}
\begin{verbatim}
> x1 <- seq(0, 1, length.out = 30)
> p1 <- predict(m1, newdata = data.frame(u = x1), interval = "credible")
> p2 <- predict(m1, newdata = data.frame(u = x1), interval = "prediction") 
\end{verbatim}
where the first argument in function \code{predict} is a fitted \code{mvrm} model, the second one is a data frame 
containing the feature vectors at which predictions are to be obtained and the last one defines the 
type of interval to be created. We applied the \code{predict} function to the two simulated datasets.   
To each of those datasets we fitted two models: the first one is the one we saw earlier, where both the mean and variance 
are modeled in terms of covariates, while the second one ignores the dependence of the variance on the covariate. 
The latter model is specified in R using
\begin{verbatim}
> model <- y ~ sm(u, k = 20, bs = "rd") | 1 
\end{verbatim}

\begin{figure}
\begin{center}
\begin{tabular}{cc}
\includegraphics[width=0.30\textwidth]{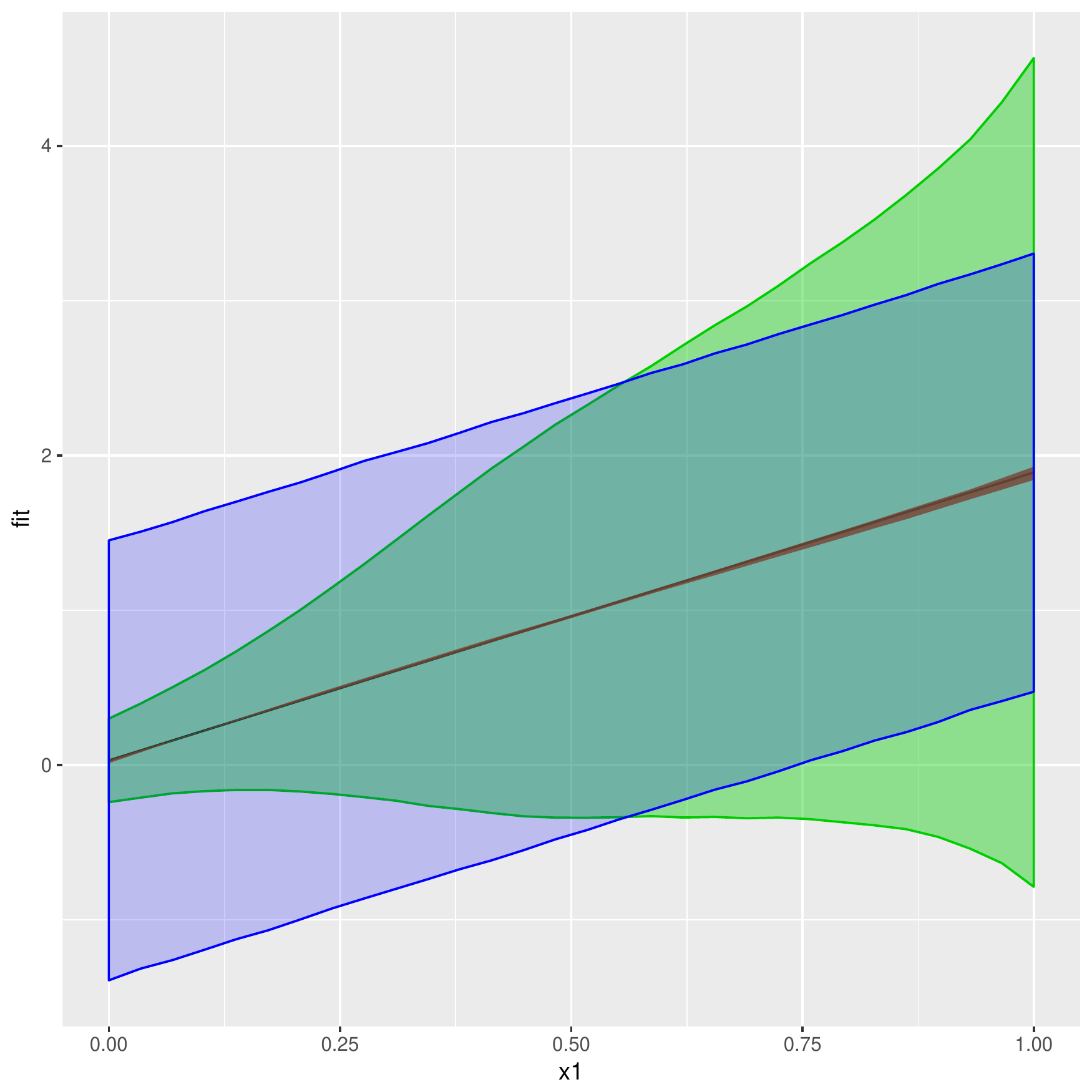} &  
\includegraphics[width=0.30\textwidth]{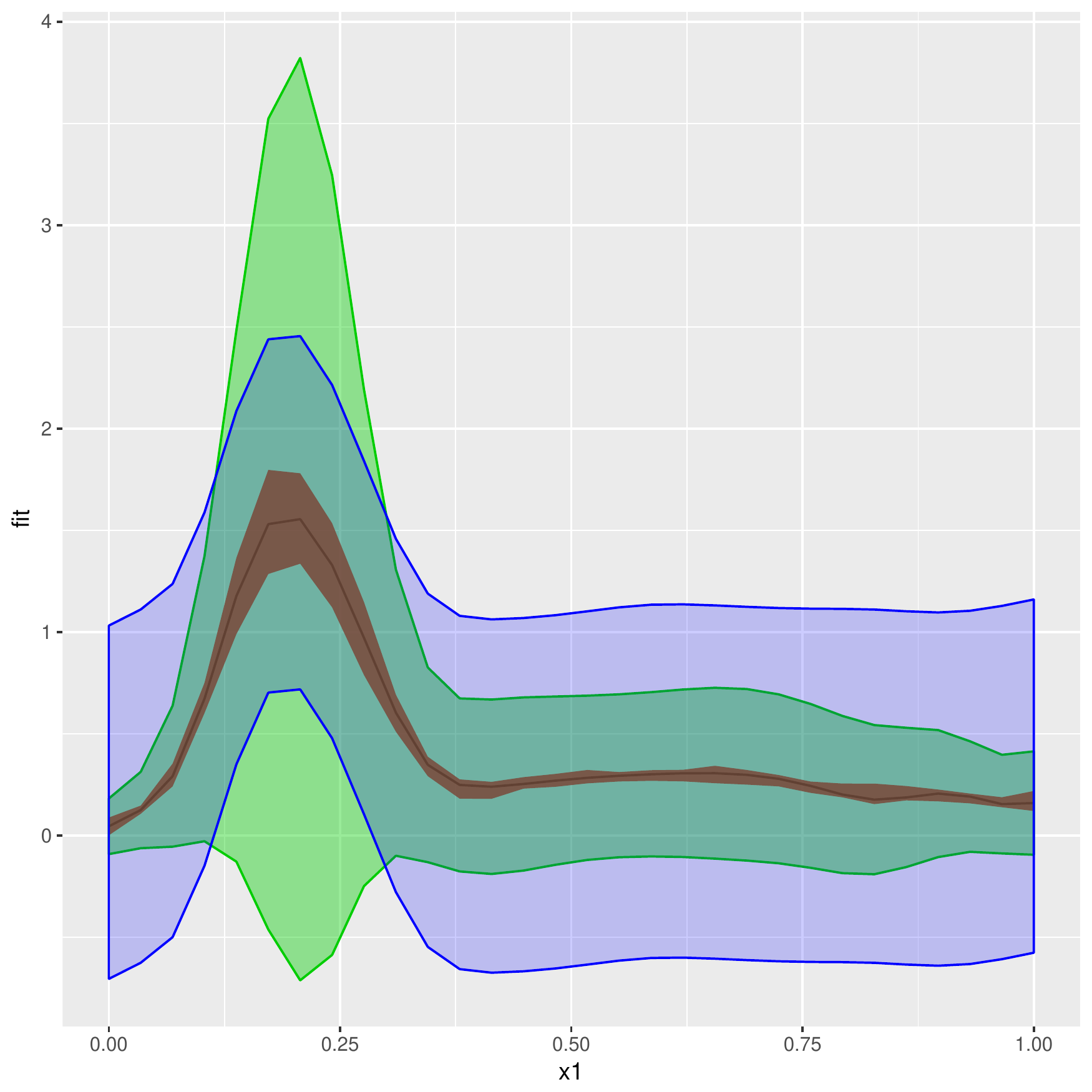} \\
(a) & (b) \\
\end{tabular}
\end{center}
\caption{Predictions results from the first two simulated datasets.
Each panel displays a credible interval and two prediction intervals, one obtained using a model that 
recognizes the dependence of the variance on the covariate and one that ignores it.}\label{sim1pred}
\end{figure}

Results are displayed in Figure~\ref{sim1pred}. Each of the two figures displays a credible interval and two
prediction intervals. The figure emphasizes a point that was discussed in the introductory 
section, that modeling the variance in terms of covariates can result in more realistic prediction intervals.
The same point was recently discussed by \citet{Mayr12}.

\subsubsection{Bivariate covariate case}

Interactions between two predictors can be modeled by appropriate specification of either the built-in \code{sm} function
or the smooth constructors from \pkg{mgcv}. Function \code{sm} can take up to two covariates, both of which may be continuous 
or one continuous and one discrete. Next we consider an example that involves two continuous covariates. An example 
involving a continuous and a discrete covariate is shown later on, in the second application to a real dataset.

Let $\uu=(u_1,u_2)^{\top}$ denote a bivariate predictor. The data-generating mechanism that we consider is 
\begin{eqnarray}
&& y(\uu) \sim N\{\mu(\uu),\sigma^2(\uu)\}, \nonumber\\
&& \mu(\uu) = 0.1 + N\left(\uu, \umu_1, \uSigma_1 \right) + N\left(\uu, \umu_2, \uSigma_2\right), \nonumber\\
&& \sigma^2(\uu) = 0.1 + \left\{N\left(\uu, \umu_1, \uSigma_1 \right) + N\left(\uu, \umu_2, \uSigma_2\right)\right\}/2, \nonumber\\
&& \umu_1 = \begin{pmatrix} 0.25 \\ 0.75 \end{pmatrix}, 
\uSigma_1 = \begin{pmatrix} 0.03 & 0.01 \\ 0.01 & 0.03 \end{pmatrix},
\umu_2 = \begin{pmatrix} 0.65 \\ 0.35 \end{pmatrix},
\uSigma_2 = \begin{pmatrix} 0.09 & 0.01 \\ 0.01 & 0.09 \end{pmatrix}.\nonumber
\end{eqnarray}
As before, $u_1$ and $u_2$ are obtained independently from uniform distributions on the unit interval. Further, 
the sample size is set to $n=500$.

In R, we simulate data from the above mechanism using
\begin{verbatim}
> mu1 <- matrix(c(0.25, 0.75))
> sigma1 <- matrix(c(0.03, 0.01, 0.01, 0.03), 2, 2)
> mu2 <- matrix(c(0.65, 0.35))
> sigma2 <- matrix(c(0.09, 0.01, 0.01, 0.09), 2, 2)
> mu <- function(x1, x2) {x <- cbind(x1, x2); 
+       0.1 + dmvnorm(x, mu1, sigma1) + dmvnorm(x, mu2, sigma2)}
> Sigma <- function(x1, x2) {x <- cbind(x1, x2); 
+          0.1 + (dmvnorm(x, mu1, sigma1) + dmvnorm(x, mu2, sigma2)) / 2}
> set.seed(1)
> n <- 500
> w1 <- runif(n)
> w2 <- runif(n)
> y <- vector()
> for (i in 1 : n) y[i] <- rnorm(1, mean = mu(w1[i], w2[i]), 
+                              sd = sqrt(Sigma(w1[i], w2[i])))
> data <- data.frame(y, w1, w2)
\end{verbatim}

We fit a model with mean and variance functions specified as 
\begin{eqnarray}
\mu(\uu) = \beta_0 + \sum_{j_1=1}^{12} \sum_{j_2=1}^{12} \beta_{j_1,j_2} \phi_{1j_1,j_2}(\uu), \quad 
\log(\sigma^2(\uu)) = \alpha_0 + \sum_{j_1=1}^{12} \sum_{j_2=1}^{12} \alpha_{j_1,j_2} \phi_{2j_1,j_2}(\uu). \nonumber\label{modelsim2}
\end{eqnarray}
The R code that fits the above model is 
\begin{verbatim}
> Model <- y ~ sm(w1, w2, k = 10, bs = "rd") | sm(w1, w2, k = 10, bs = "rd")
> m2 <- mvrm(formula = Model, data = data, sweeps = 10000, burn = 5000, thin = 2, 
+            seed = 1, StorageDir = DIR)
\end{verbatim}

\begin{figure}
\begin{center}
\begin{tabular}{cc}
\includegraphics[width=0.45\textwidth]{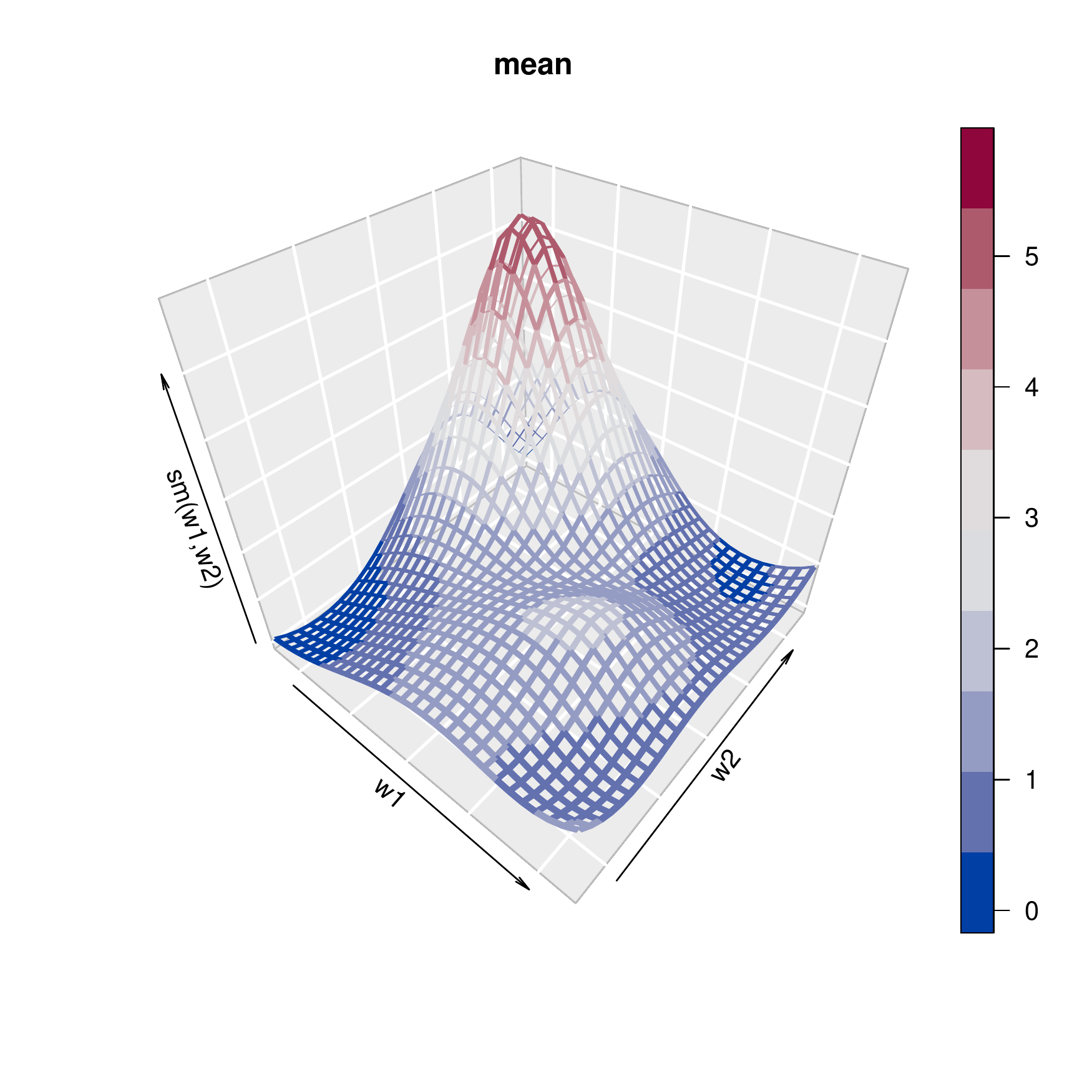} &  
\includegraphics[width=0.45\textwidth]{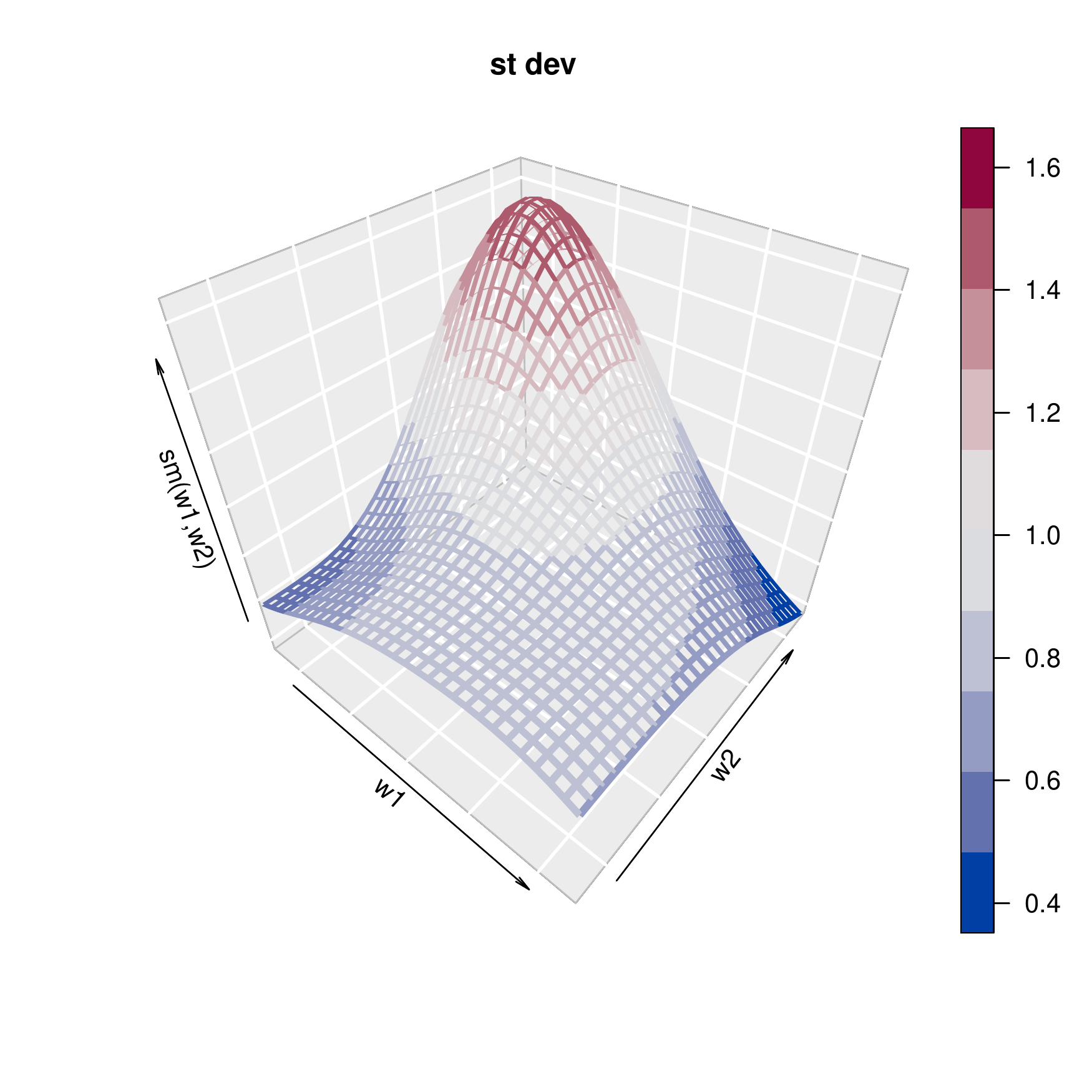} \\
(a) & (b) \\
\end{tabular}
\end{center}
\caption{Bivariate simulation study results with two continuous covariates. Posterior means of the (a) mean and (b) 
standard deviation function.}\label{sim2}
\end{figure}

As in the univariate case, convergence assessment and univariate posterior summaries may be obtained by using
function \code{mvrm2mcmc} in conjunction with functions \code{plot.mcmc} and \code{summary.mcmc}. Further, 
summaries of the `mvrm' fits may be obtained using functions \code{print.mvrm} and \code{summary.mvrm}. 
Plots of the bivariate effects may be obtained using function \code{plot.mvrm}. This is shown below,
where argument \code{plotOptions} utilizes package \CRANpkg{colorspace} \citep{ZHM09}. 
\begin{verbatim}
> plot(x = m2, model = "mean", term = "sm(w1,w2)", static = TRUE, 
+      plotOptions = list(col = diverge_hcl(n = 10)))
> plot(x = m2, model = "stdev", term = "sm(w1,w2)", static = TRUE, 
+      plotOptions = list(col = diverge_hcl(n = 10)))
\end{verbatim}
Results are shown in Figure~\ref{sim2}. For bivariate predictors, function \code{plot.mvrm}
calls function \code{ribbon3D} from package \pkg{plot3D}. 
Dynamic plots, viewable in a browser, can be created by replacing the default \code{static = TRUE} by 
\code{static = FALSE}. When the latter option is specified, function \code{plot.mvrm} calls function 
\code{scatterplot3js} from package \pkg{threejs}. Users may pass their own options to 
\code{plot.mvrm} via the \code{plotOptions} argument.  

\subsubsection{Multiple covariate case}

We consider fitting general additive models for the mean and variance functions in a simulated example
with four independent continuous covariates. In this scenario, we set $n=1000$. Further the covariates
$\uw=(w_1,w_2,w_3,w_4)^{\top}$
are simulated independently from a uniform distribution on the unit interval. The data-generating mechanism that we consider
is 
\begin{eqnarray}
&& Y(\uw) \sim N(\mu(\uw),\sigma^2(\uw)), \nonumber \\
&& \mu(\uw) = \sum_{j=1}^4 \mu_j(w_j) \quad \text{and} \quad \sigma(\uw) = \prod_{j=1}^4 \sigma_j(w_j) \nonumber
\end{eqnarray}
where functions $\mu_j, \sigma_j, j=1,\dots,4,$ are specified below
\begin{enumerate}
\item $\mu_1(w_1) = 1.5 w_1, \sigma_1(w_1)=\left\{N(w_1,\mu=0.2,\sigma^2=0.004)+N(w_1,\mu=0.6,\sigma^2=0.1)\right\}/2$,
\item $\mu_2(w_2) = \left\{N(w_2,\mu=0.2,\sigma^2=0.004)+N(w_2,\mu=0.6,\sigma^2=0.1)\right\}/2,$\\ 
      $\sigma_2(w_2)=0.6+0.5\sin(2 \pi w_2)$,
\item $\mu_3(w_3) = 1+ \sin(2 \pi w_3), \sigma_3(w_3)=1.1-w_3$,
\item $\mu_4(w_4) = -w_4, \sigma_4(w_4)=0.2+1.5w_4$.
\end{enumerate}
 
To the generated dataset we fit a model with mean and variance functions modeled as 
\begin{eqnarray}
\mu(\uw) = \beta_0 + \sum_{k=1}^{4} \sum_{j=1}^{16} \beta_{kj} \phi_{kj}(w_k) \quad \text{and} \quad  
\log\{\sigma^2(\uw)\} = \alpha_0 + \sum_{k=1}^{4} \sum_{j=1}^{16} \alpha_{kj} \phi_{kj}(w_k). \nonumber\label{modelsim3}
\end{eqnarray}
Fitting the above model to the simulated data is achieved by the following R code
\begin{verbatim}
> Model <- y ~ sm(w1, k = 15, bs = "rd") + sm(w2, k = 15, bs = "rd") + 
+              sm(w3, k = 15, bs = "rd") + sm(w4, k = 15, bs = "rd") |
+              sm(w1, k = 15, bs = "rd") + sm(w2, k = 15, bs = "rd") + 
+              sm(w3, k = 15, bs = "rd") + sm(w4, k = 15, bs = "rd")
> m3 <- mvrm(formula = Model, data = data, sweeps = 50000, burn = 25000, 
+            thin = 5, seed = 1, StorageDir = DIR)
\end{verbatim}
By default function \code{sm} utilizes the radial basis functions, hence there is no need to specify \code{bs = "rd"}
as we did earlier, if radial basis functions are preferred over thin plate splines. Further, we have selected 
\code{k = 15} for all smooth functions. However, there is no restriction to the
number of knots and certainly one can select a different number of knots for each smooth function. 

As discussed previously, for each term that appears in the right-hand side of the mean and
variance functions, the model incorporates indicator variables that specify which basis functions are to be included 
and which are to be excluded from the model. For the current example, the indicator variables are denoted by 
$\gamma_{kj}$ and $\delta_{kj}, k=1,2,3,4,j=1,\dots,16$. 
The prior probabilities that variables are included were specified in (\ref{gam.priors}) and they are specific to 
each term,  
$\pi_{\mu_k} \sim \text{Beta}(c_{\mu_k},d_{\mu_k}), \quad
\pi_{\sigma_k} \sim \text{Beta}(c_{\sigma_k},d_{\sigma_k}), k=1,2,3,4.$
The default option \code{pi.muPrior = "Beta(1,1)"} specifies that $\pi_{\mu_k} \sim \text{Beta}(1,1),k=1,2,3,4.$
Further, by setting, for example, \code{pi.muPrior = "Beta(2,1)"} we specify that $\pi_{\mu_k} \sim \text{Beta}(2,1),k=1,2,3,4.$
To specify a different Beta prior for each of the four terms, \code{pi.muPrior} will have to be specified as a vector 
of length four, as an example, \code{pi.muPrior = c("Beta(1,1)","Beta(2,1)","Beta(3,1)","Beta(4,1)")}.
Specification of the priors for $\pi_{\sigma_k}$ is done in a similar way, via argument \code{pi.sigmaPrior}.  

\begin{figure}
\begin{center}
\begin{tabular}{cc}
\includegraphics[width=0.43\textwidth,height=0.20\textheight]{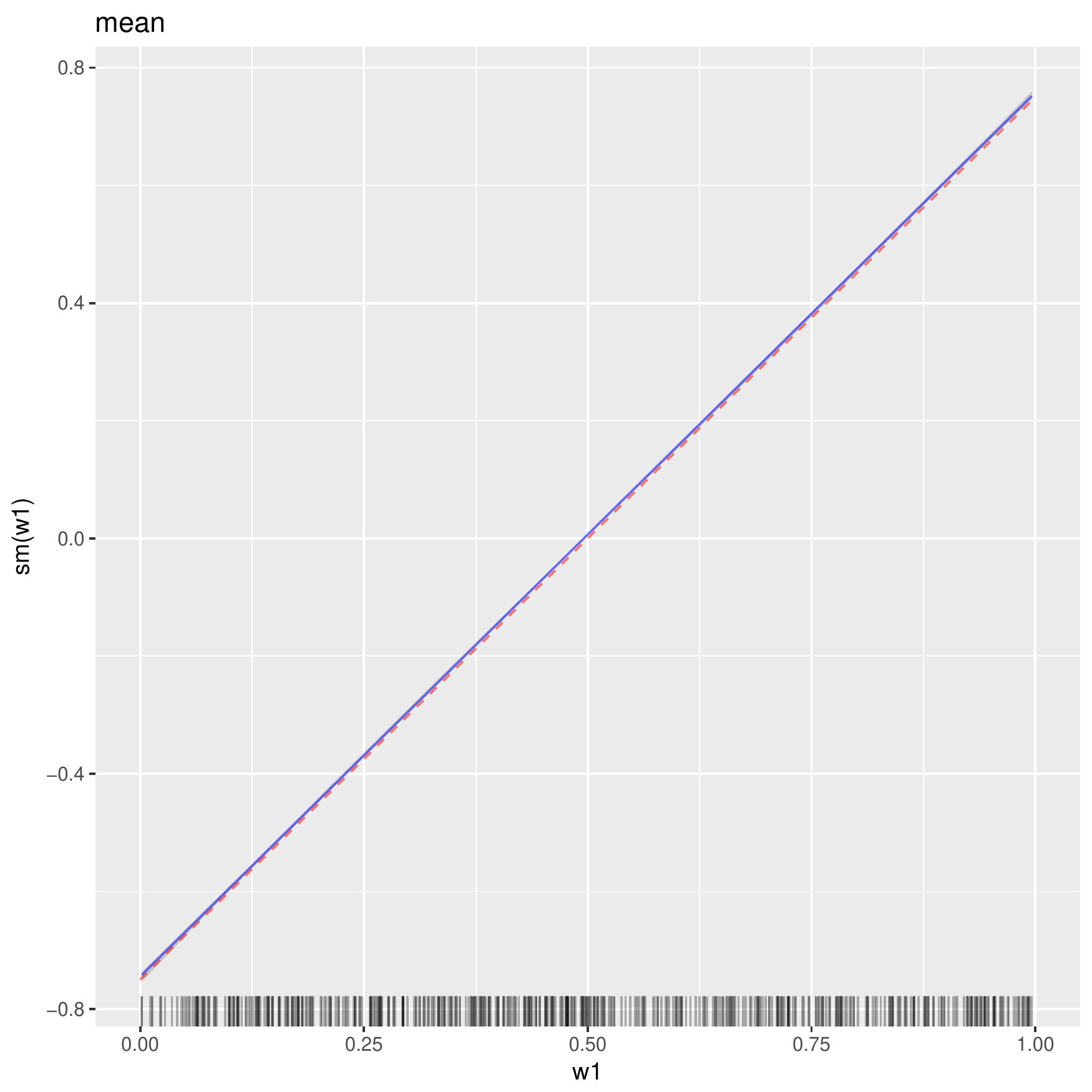} &  
\includegraphics[width=0.43\textwidth,height=0.20\textheight]{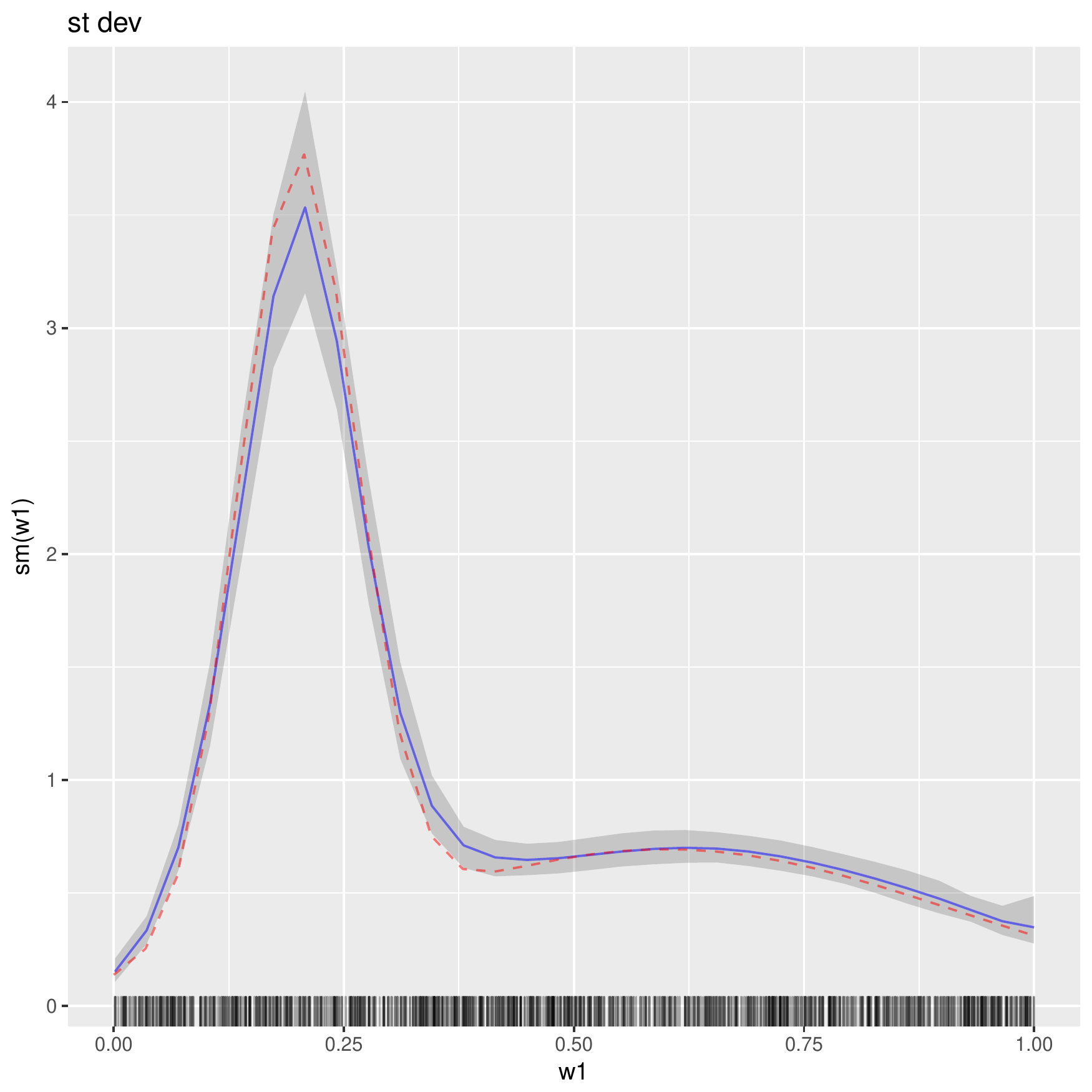} \\
\includegraphics[width=0.43\textwidth,height=0.20\textheight]{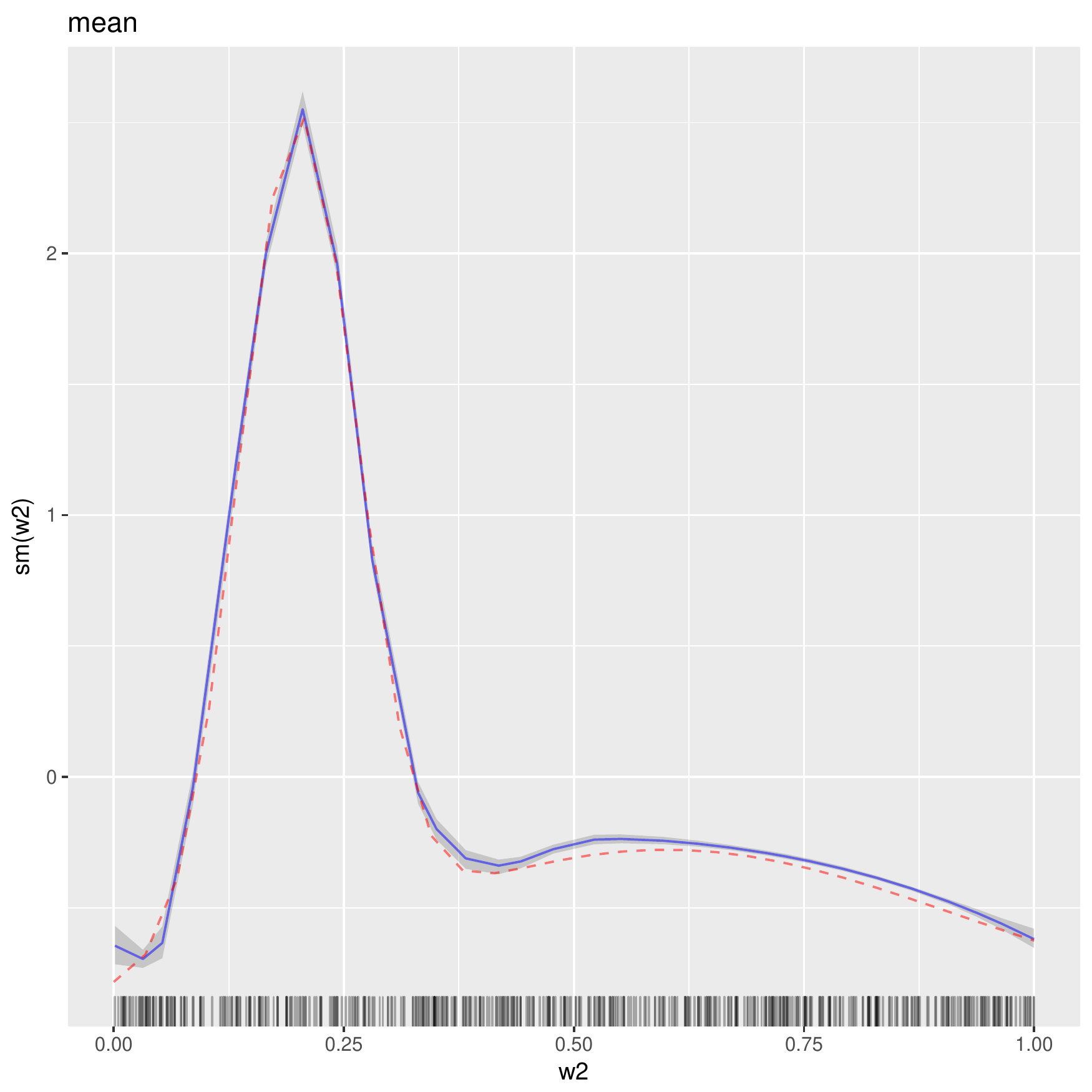} &  
\includegraphics[width=0.43\textwidth,height=0.20\textheight]{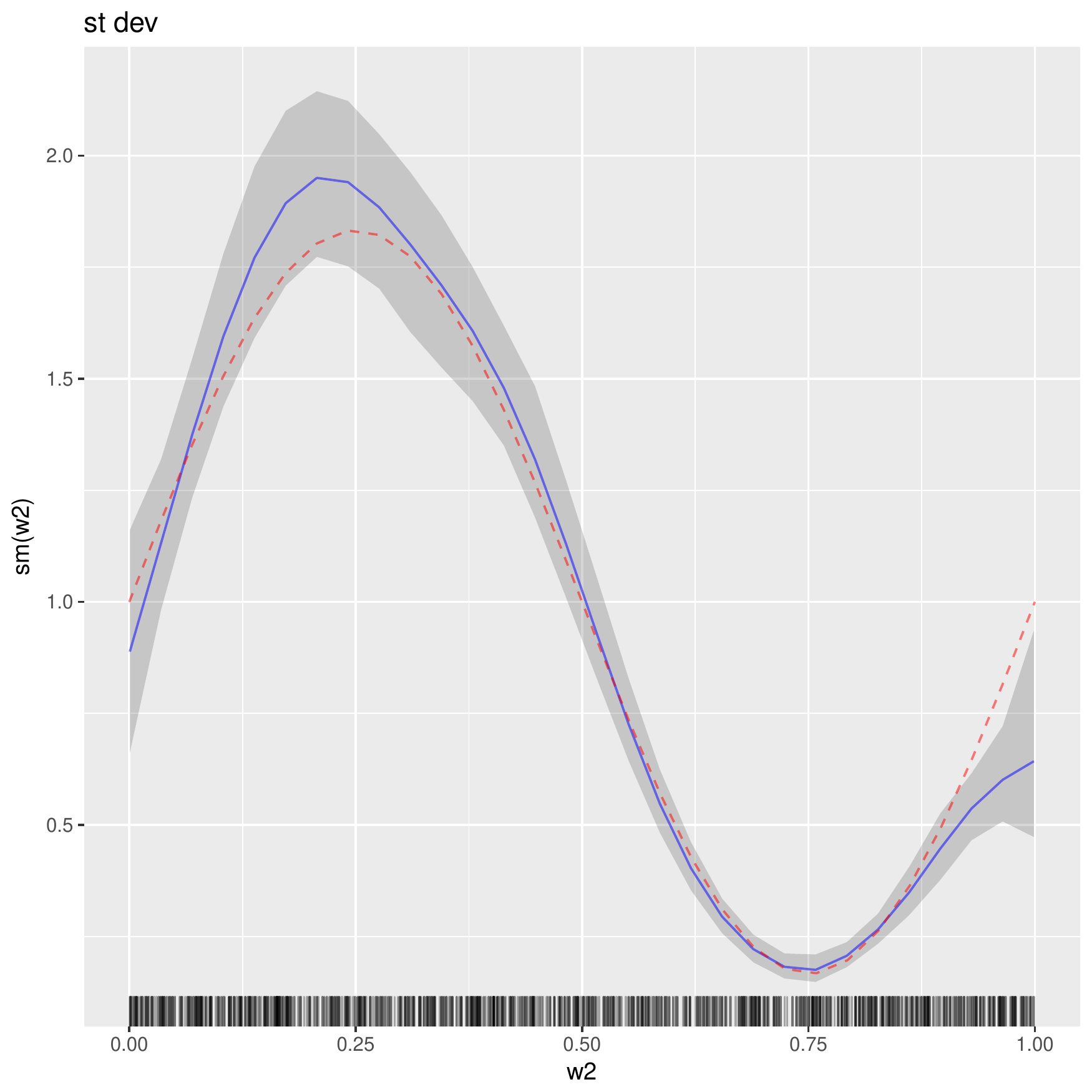} \\
\includegraphics[width=0.43\textwidth,height=0.20\textheight]{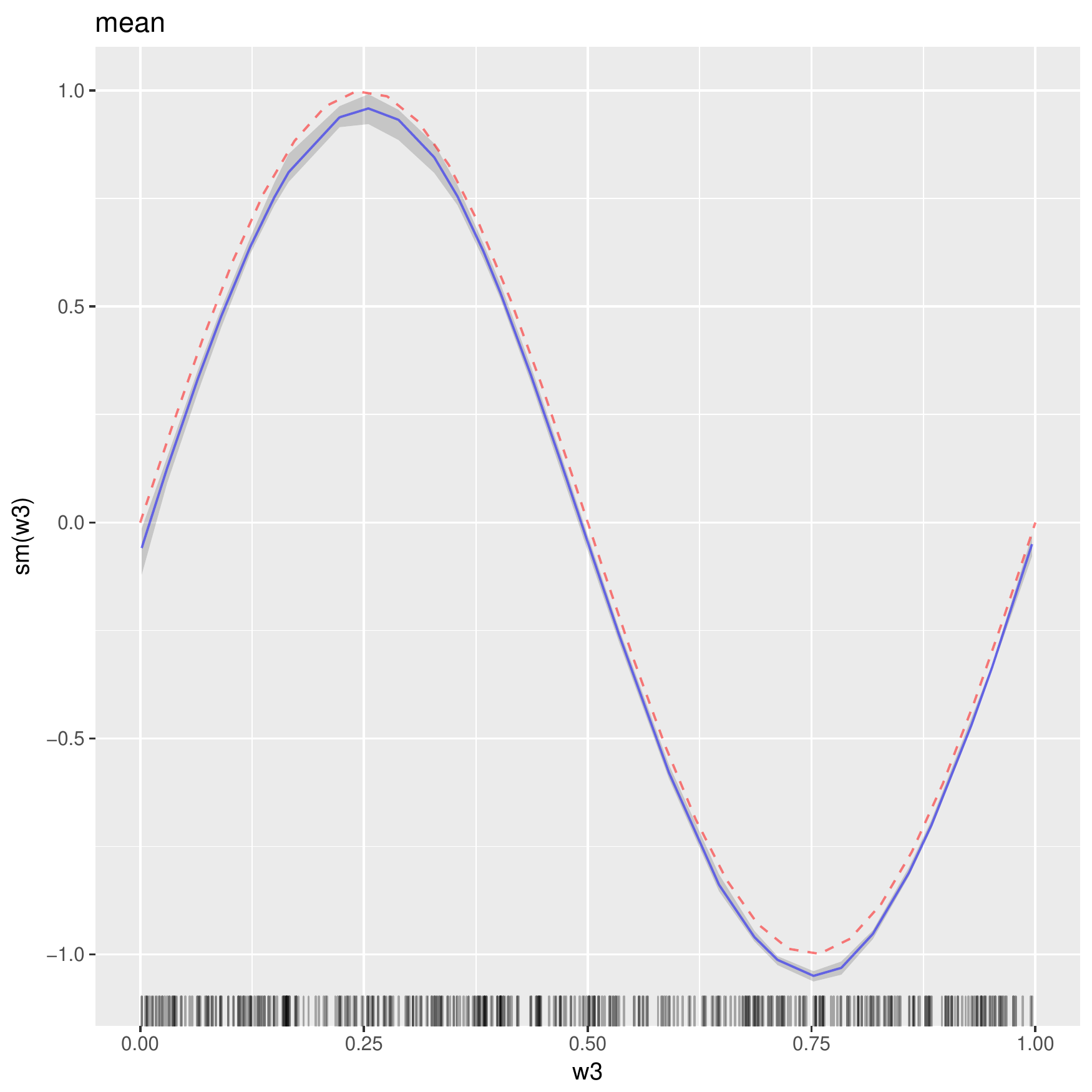} &  
\includegraphics[width=0.43\textwidth,height=0.20\textheight]{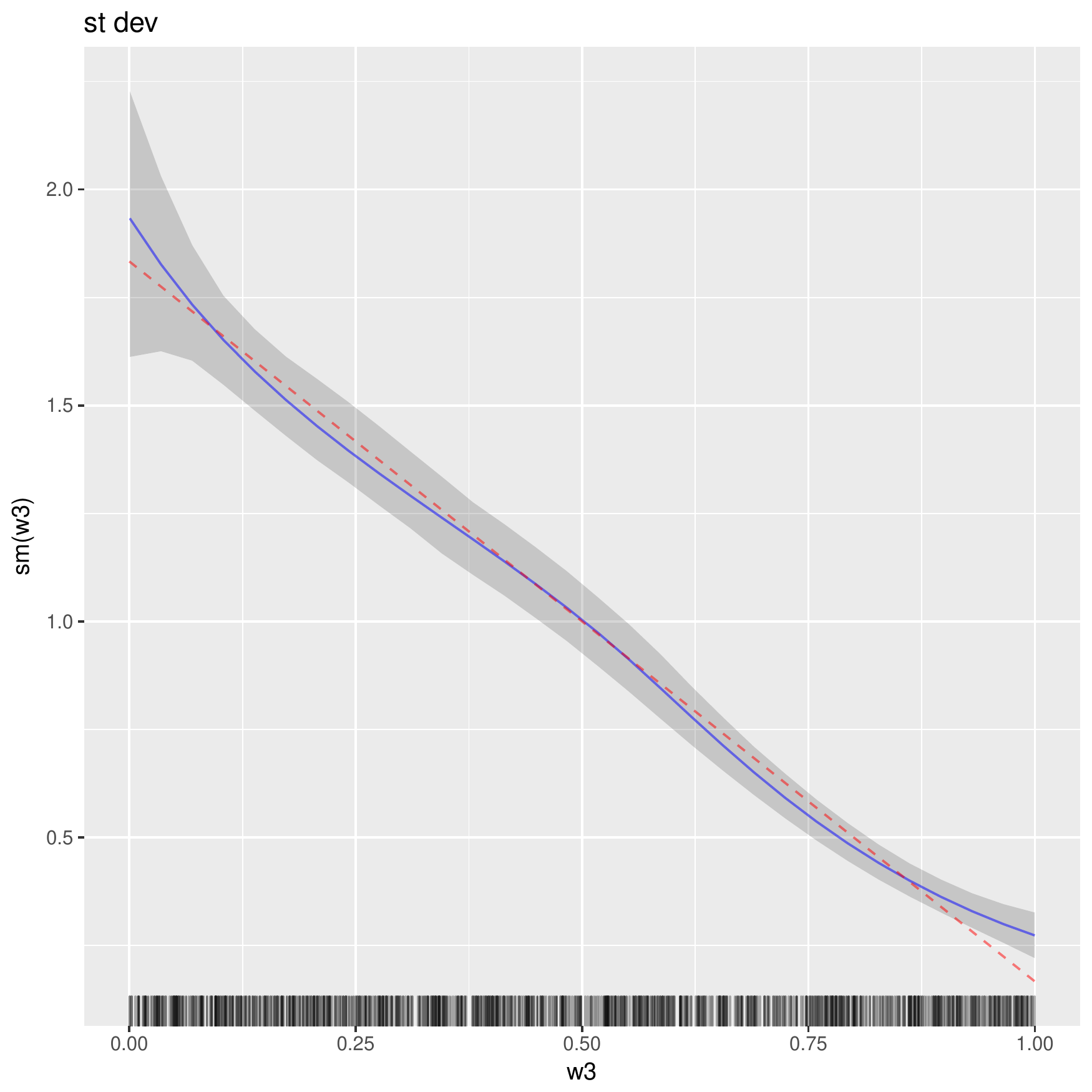} \\
\includegraphics[width=0.43\textwidth,height=0.20\textheight]{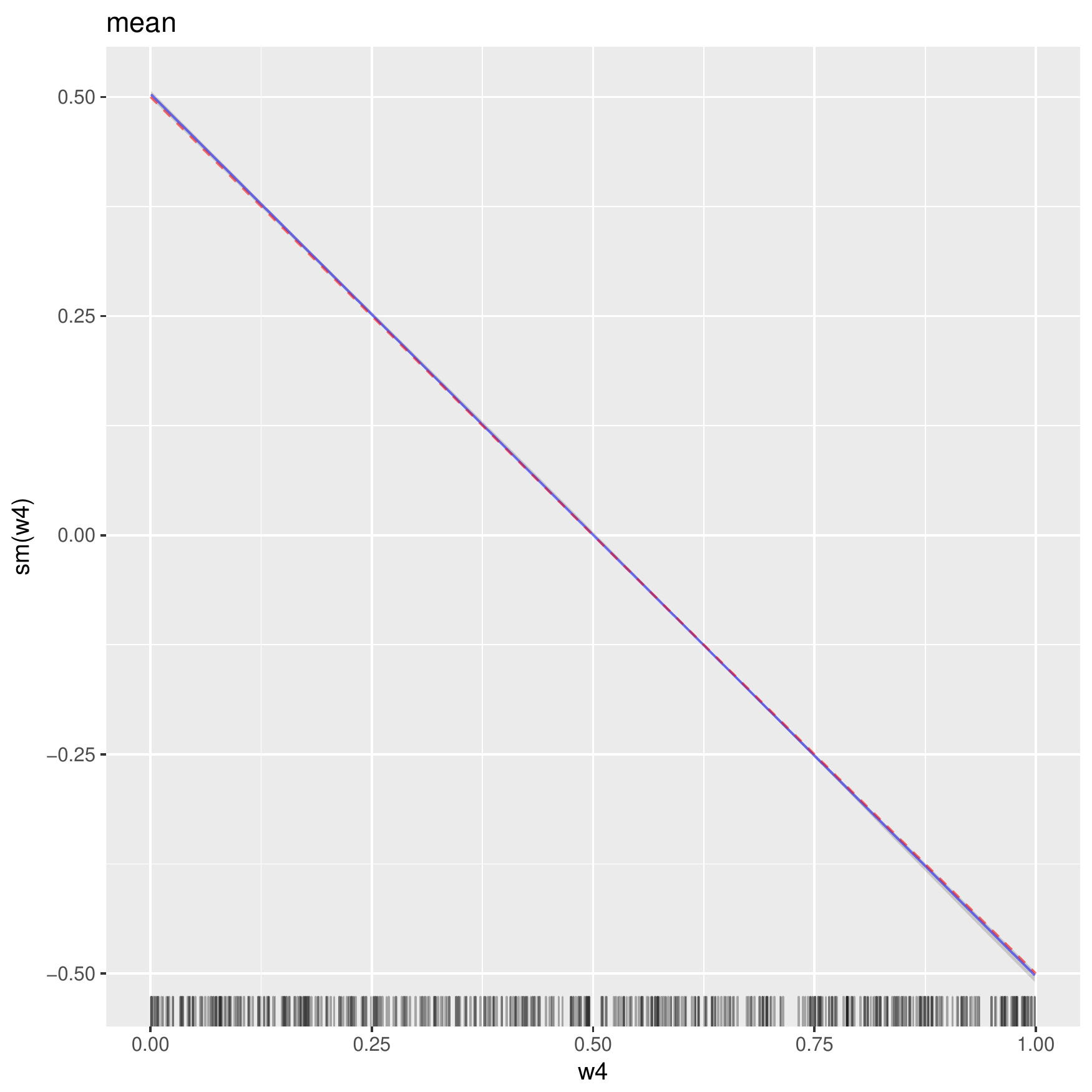} &  
\includegraphics[width=0.43\textwidth,height=0.20\textheight]{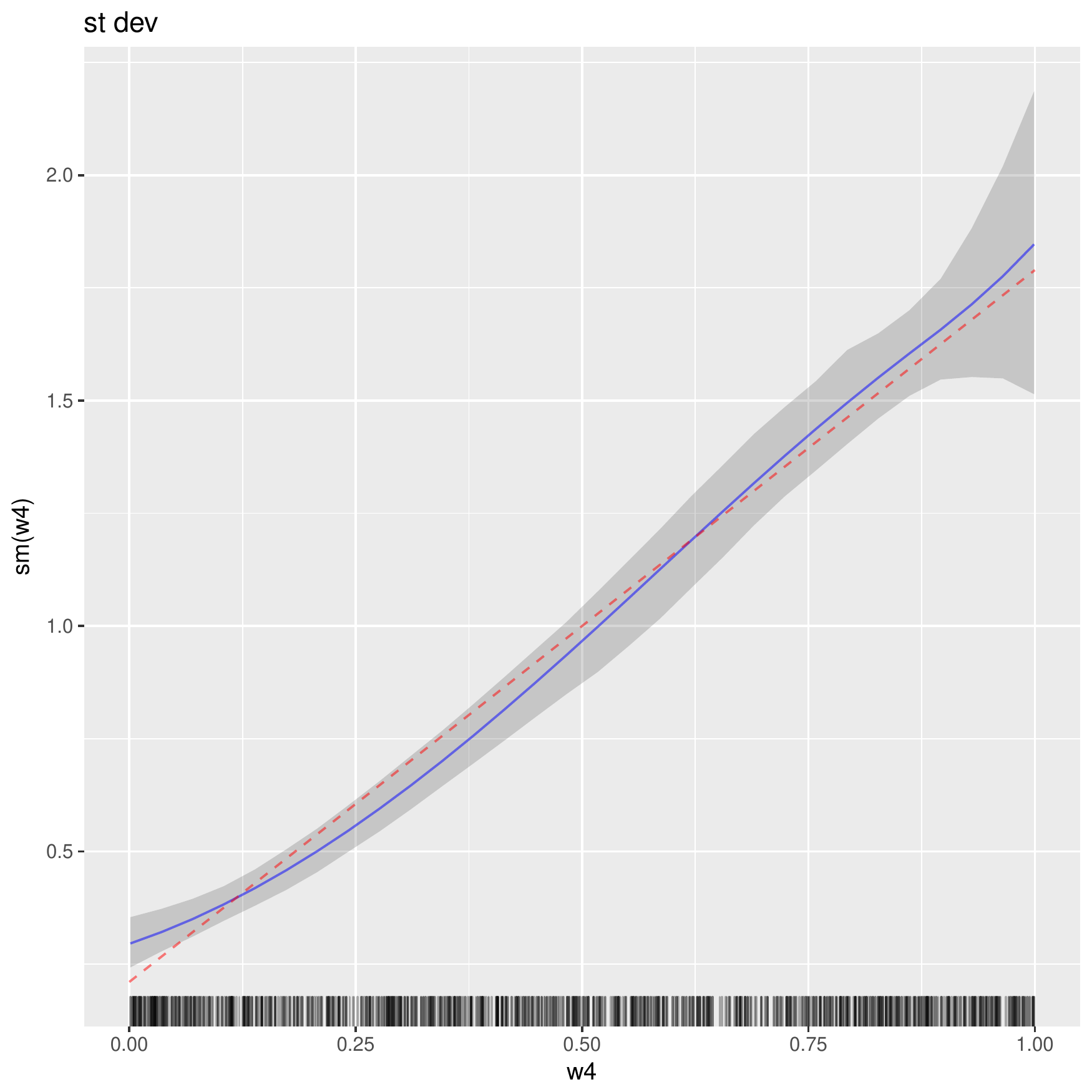} \\
\end{tabular}
\end{center}
\caption{Multiple covariate simulation study results. The column on the left-hand side presents 
the true and estimated mean functions, along with $99\%$ credible intervals. The column on the 
right-hand side presents the true and estimated standard deviation functions, along with $95\%$ credible intervals. 
In all panels, the
truth is represented by dashed (red color) lines, the estimated functions by solid (blue color) lines, and the
credible intervals by gray color. }\label{gamsim}
\end{figure}

We conclude this section by presenting plots of the four terms in the mean and variance models. The plots are 
presented in Figure~\ref{gamsim}. We provide a few details on how function \code{plot} works in the presence of multiple terms, 
and how the comparison between true and estimated effects is made. Starting with the mean function, to create the 
relevant plots, that appear on the left panels of Figure~\ref{gamsim}, function \code{plot} considers only the part of 
the mean function $\mu(\uu)$ that is related to the chosen \code{term} while leaving all other terms out. 
For instance, in the code below we choose \code{term = "sm(u1)"} and hence we plot the posterior 
mean and a posterior credible interval for $\sum_{j=1}^{16} \beta_{1j} \phi_{1j}(u_1)$, where the intercept $\beta_0$ is 
left out by option \code{intercept = FALSE}. Further, comparison is made with a centered version of the true curve, 
represented by the dashed (red color) line and obtained by the first three lines of code below.   
\begin{verbatim}
> x1 <- seq(0, 1, length.out = 30)
> y1 <- mu1(x1)
> y1 <- y1 - mean(y1)
> PlotOptions <- list(geom_line(aes_string(x = x1, y = y1), 
+                     col = 2, alpha = 0.5, lty = 2))
> plot(x = m3, model = "mean", term = "sm(w1)", plotOptions = PlotOptions, 
+      intercept = FALSE, centreEffects = FALSE, quantiles = c(0.005, 1 - 0.005)) 
\end{verbatim}

The plots of the four standard deviation terms are shown in the right panels of Figure~\ref{gamsim}. Again, these are 
created by considering only the part of the model for $\sigma(\uu)$ that is related to the chosen \code{term}. 
For instance, below we choose \code{term = "sm(u1)"}. Hence, in this case the plot will present the posterior mean 
and a posterior credible interval for $\exp\{\sum_{j=1}^{16} \alpha_{1j} \phi_{1j}(u_1)/2\}$, where the intercept $\alpha_0$ 
is left out by option \code{intercept = FALSE}.
Option \code{centreEffects = TRUE} scales the posterior realizations of $\exp\{\sum_{j=1}^{16} \alpha_{1j} \phi_{1j}(u_1)/2\}$
before plotting them, where the scaling is done in such a way that the realized function has mean one over the range of the 
predictor. Further, the comparison is made with a scaled version of the true curve, where again the scaling is done  
to achieve mean one. This is shown below and it is in the 
spirit of \citet{Chan06} who discuss the differences between the data generating mechanism and the fitted model. 
\begin{verbatim}
> y1 <- stdev1(x1) / mean(stdev1(x1))
> PlotOptions <- list(geom_line(aes_string(x = x1, y = y1), 
+                     col = 2, alpha = 0.5, lty = 2))
> plot(x = m3, model = "stdev", term = "sm(w1)", plotOptions = PlotOptions, 
+      intercept = FALSE, centreEffects = TRUE, quantiles = c(0.025, 1 - 0.025))
\end{verbatim}

\subsection{Data analyses} 

In this section we present four empirical applications. 

\subsubsection{Wage and age}

In the first empirical application, we analyse a dataset from \citet{Pagan} that is available in the R package \CRANpkg{np} \citep{np}. 
The dataset consists of $n=205$ observations on dependent variable \code{logwage},  
the logarithm of the individual's wage, and covariate \code{age}, the individual's age.  
The dataset comes from the 1971 Census of Canada Public Use Sample Tapes and the sampling units it involves 
are males of common education. Hence, the investigation of the relationship between age and the logarithm of wage  
is carried out controlling for the two potentially important covariates education and gender. 

We utilize the following R code to specify flexible models for the mean and variance functions, and to 
obtain $5,000$ posterior samples, after a burn-in period of $25,000$ samples and a thinning period of $5$.  
\begin{verbatim}
> data(cps71)
> DIR <- getwd()
> model <- logwage ~ sm(age, k = 30, bs = "rd") | sm(age, k = 30, bs = "rd")
> m4 <- mvrm(formula = model, data = cps71, sweeps = 50000, 
+            burn = 25000, thin = 5, seed = 1, StorageDir = DIR)
\end{verbatim}

\begin{figure}
\begin{center}
\begin{tabular}{cc}
\includegraphics[width=0.45\textwidth, height=0.2\textheight]{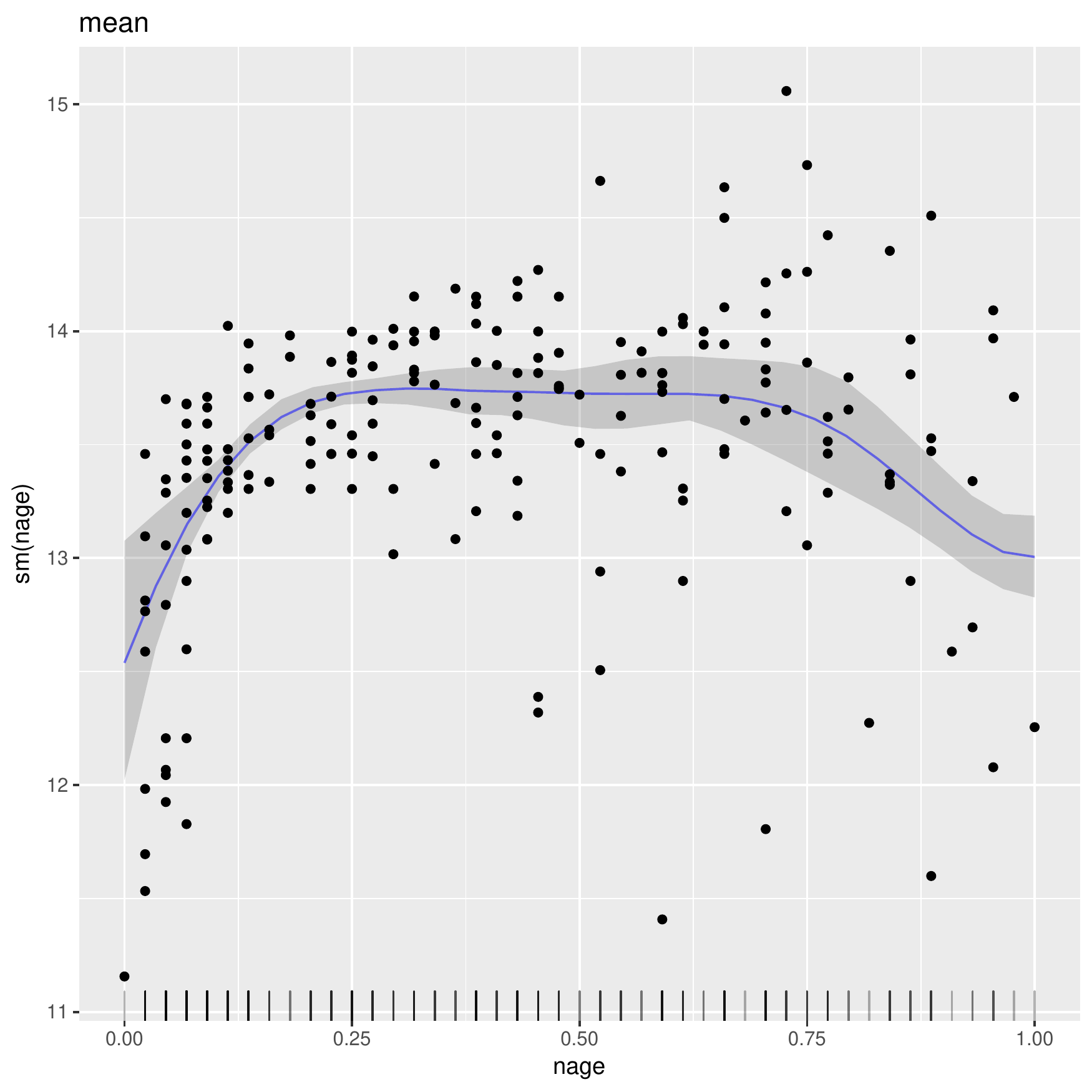} &  
\includegraphics[width=0.45\textwidth, height=0.2\textheight]{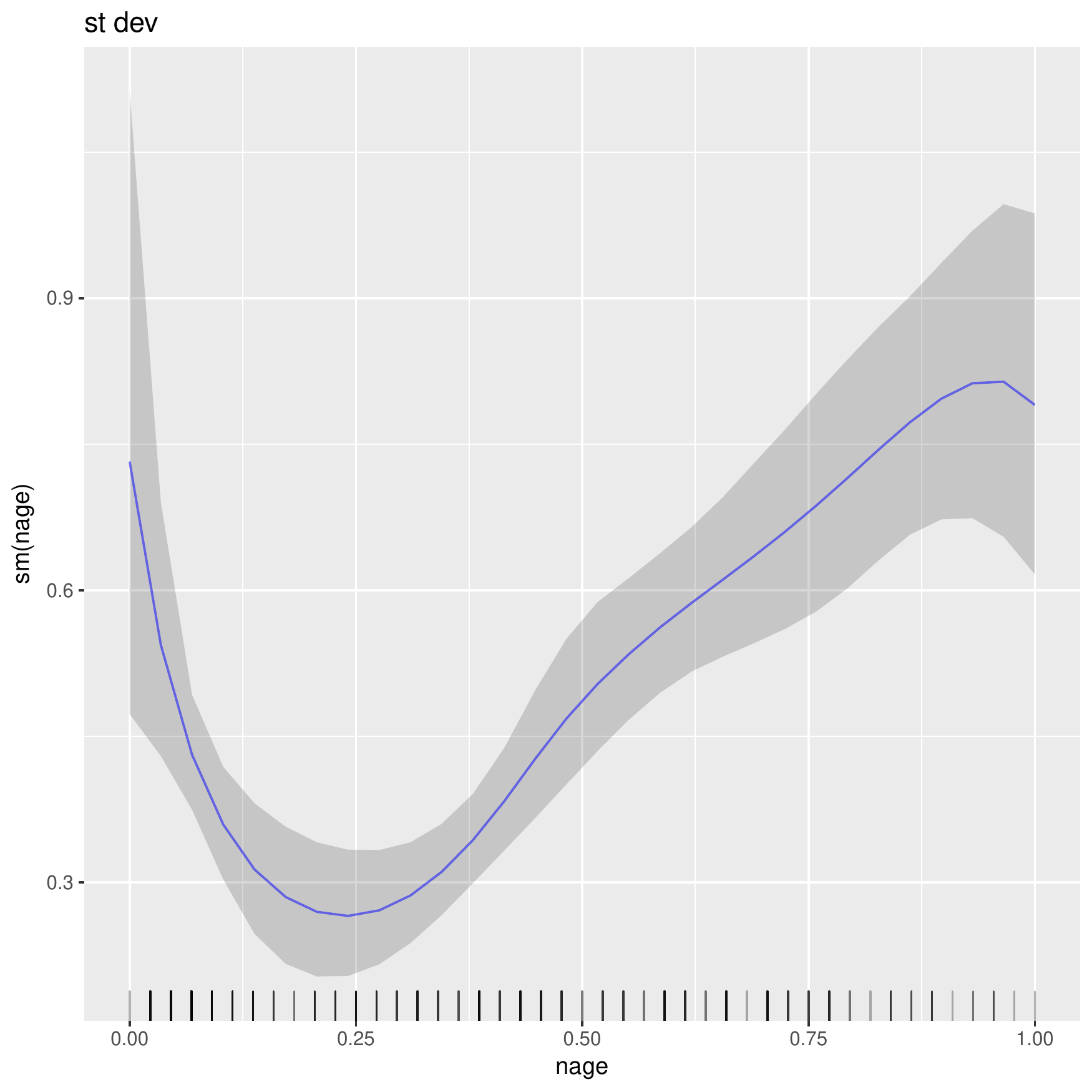} \\
(a) & (b) \\
\end{tabular}
\end{center}
\caption{Results from the data analysis on the relationship between age and the logarithm of wage.
Panel (a) shows the posterior mean and an $80\%$ credible interval of the mean function, and the observed data-points. Panel (b)
shows the posterior mean and an $80\%$ credible interval of the standard deviation function.}\label{udaf1}
\end{figure}

After checking convergence, we use the following code to create the plots that appear in Figure~\ref{udaf1}. 
\begin{verbatim}
> wagePlotOptions <- list(geom_point(data = cps71, aes(x = age, y = logwage)))
> plot(x = m4, model = "mean", term = "sm(age)", plotOptions = wagePlotOptions)
> plot(x = m4, model = "stdev", term = "sm(age)")
\end{verbatim}

Figure~\ref{udaf1} (a) shows the posterior mean and an $80\%$ credible interval for the mean function
and it suggests a quadratic relationship between \code{age} and \code{logwage}. 
Figure~\ref{udaf1} (b) shows the posterior mean and an $80\%$ credible interval for the standard deviation 
function. It suggest a complex relationship between \code{age} and the variability in \code{logwage}.
The relationship suggested by Figure~\ref{udaf1} (b) is also suggested by the spread of the data-points around the estimated mean 
in Figure~\ref{udaf1}
(a). At ages around $20$ years the variability in \code{logwage} is high. It then reduces until about 
age $30$, to start increasing again until about age $45$. From age $45$ to $60$ it remains stable but high,
while for ages above $60$, Figure~\ref{udaf1} (b) suggests further increase in the variability, but the 
wide credible interval suggests high uncertainty over this age range. 

\subsubsection{Wage and multiple covariates}

In the second empirical application, we analyse a dataset from \citet{Wooldridge} that is also available in R package \pkg{np}.
The response variable here is the logarithm of the individual's hourly wage (\code{lwage}) while the covariates  
include the years of education (\code{educ}), the years of experience (\code{exper}), the
years with the current employer (\code{tenure}), the individual's gender (named as 
\code{female} within the dataset, with levels \code{Female} and \code{Male}), and 
marital status (named as \code{married} with levels \code{Married} and \code{Notmarried}). 
The dataset consists of $n=526$ independent observations. We analyse the first three covariates as continuous
and the last two as discrete. 

As the variance function is modeled in terms of an exponential, see (\ref{mv2}), to avoid
potential numerical problems, we transform the three continuous variables to have range in the interval $[0,1]$, using
\begin{verbatim}
> data(wage1)
> wage1$ntenure <- wage1$tenure / max(wage1$tenure)
> wage1$nexper <- wage1$exper / max(wage1$exper)
> wage1$neduc <- wage1$educ / max(wage1$educ)
\end{verbatim}

We choose to fit the following mean and variance models to the data
\begin{eqnarray}
&&\mu_i = \beta_0 + \beta_1 \;\texttt{married}_i+ f_1(\texttt{ntenure}_i) + 
          f_2(\texttt{neduc}_i) + f_3(\texttt{nexper}_i,\texttt{female}_i), \nonumber\\
&&\log(\sigma^2_i) = \alpha_0 + f_4(\texttt{nexper}_i). \nonumber
\end{eqnarray}
We note that, as it turns out, an interaction between variables \code{nexper} and \code{female} is not necessary for the 
current data analysis. However, we choose to add this term in the mean model in order to illustrate how interaction terms 
can be specified. We illustrate further options below. 
\begin{verbatim}
> knots1 <- seq(min(wage1$nexper), max(wage1$nexper), length.out = 30)
> knots2 <- c(0, 1)
> knotsD <- expand.grid(knots1, knots2)
> model <- lwage ~ fmarried + sm(ntenure) + sm(neduc, knots=data.frame(knots = 
+ seq(min(wage1$neduc), max(wage1$neduc), length.out = 15))) + 
+ sm(nexper, ffemale, knots = knotsD) | sm(nexper, knots=data.frame(knots = 
+ seq(min(wage1$nexper), max(wage1$nexper), length.out=15)))
\end{verbatim}
The first three lines of the R code above specify the matrix of (potential) knots to be used for representing
$f_3(\texttt{nexper,female})$. Knots may be left unspecified, in which case the defaults in function 
\code{sm} will be used. Furthermore, in the specification of the mean model we use \code{sm(ntenure)}. By this, we chose to 
represent $f_1$ utilizing the default $10$ knots and the radial basis functions. Further, the specification of $f_2$
in the mean model illustrates how users can specify their own knots for univariate functions. In the current example, 
we select $15$ knots uniformly spread over the range of \code{neduc}. 
Fifteen knots are also used to represent $f_4$ within the variance model. 

The following code is used to obtain samples from the posteriors of the model parameters. 
\begin{verbatim}
> DIR <- getwd()
> m5 <- mvrm(formula = model, data = wage1, sweeps = 100000, 
+            burn = 25000, thin = 5, seed = 1, StorageDir = DIR))
\end{verbatim}

\begin{figure}
\begin{center}
\begin{tabular}{cc}
\includegraphics[width=0.40\textwidth,height=0.16\textheight]{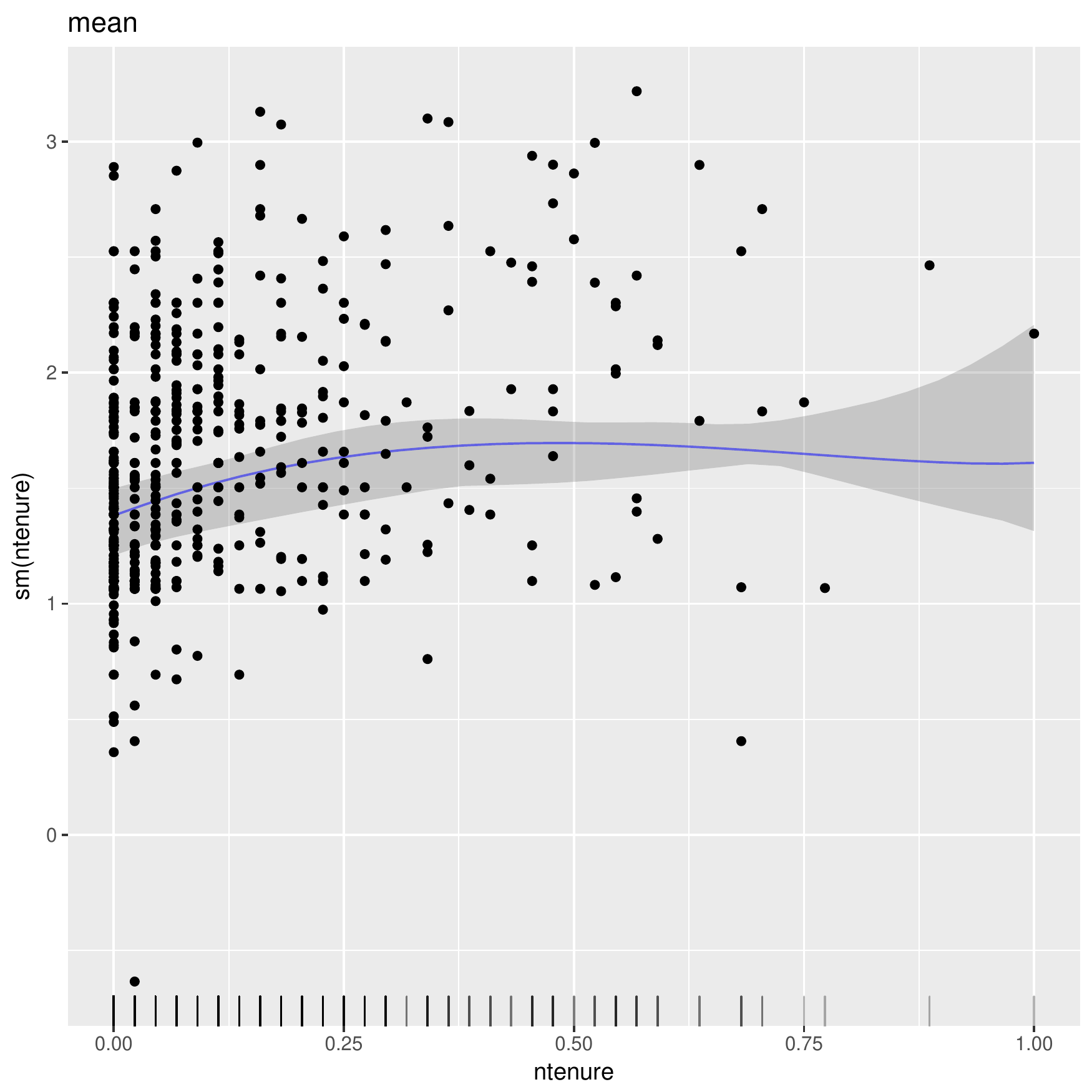} &  
\includegraphics[width=0.40\textwidth,height=0.16\textheight]{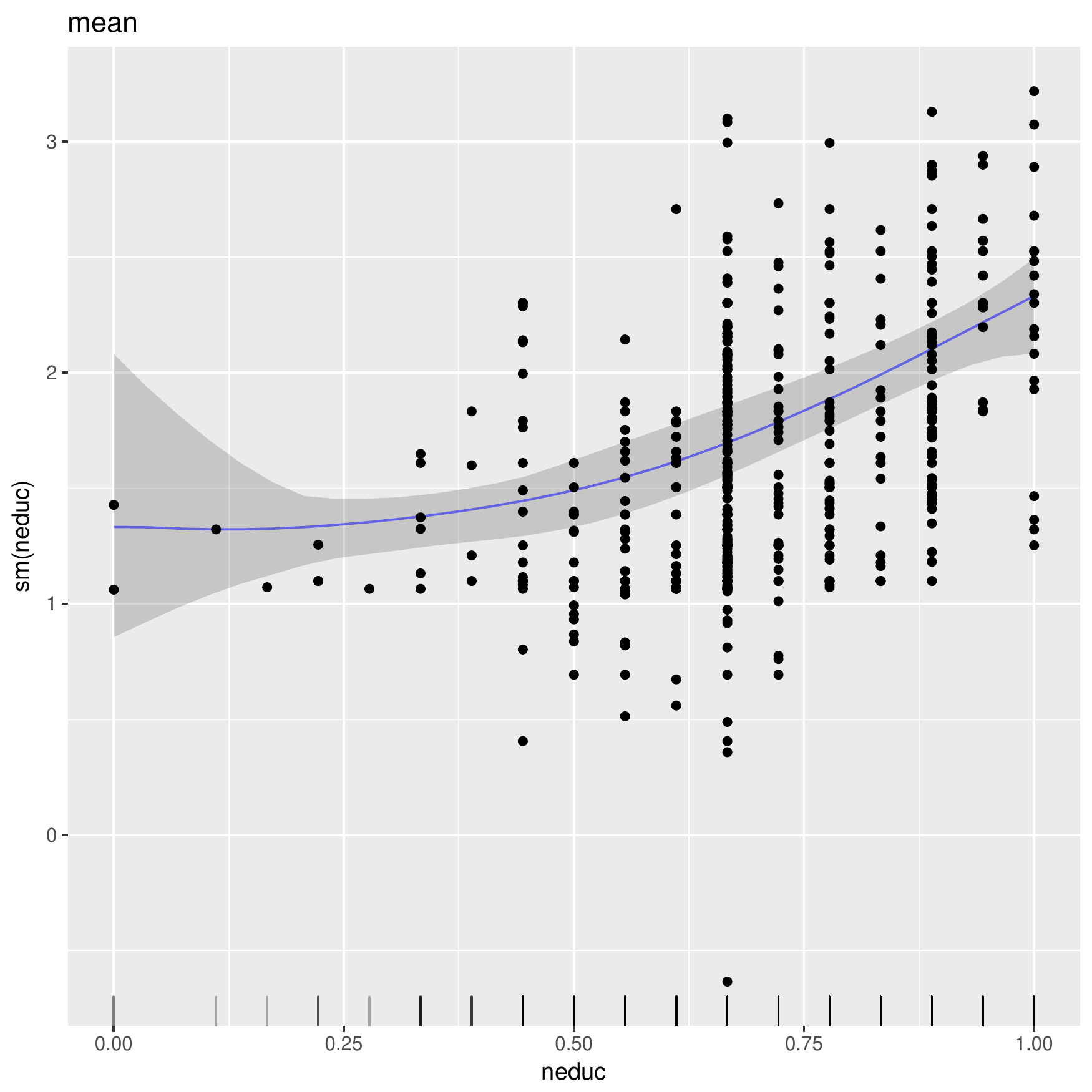} \\
(a) & (b) \\
\includegraphics[width=0.40\textwidth,height=0.16\textheight]{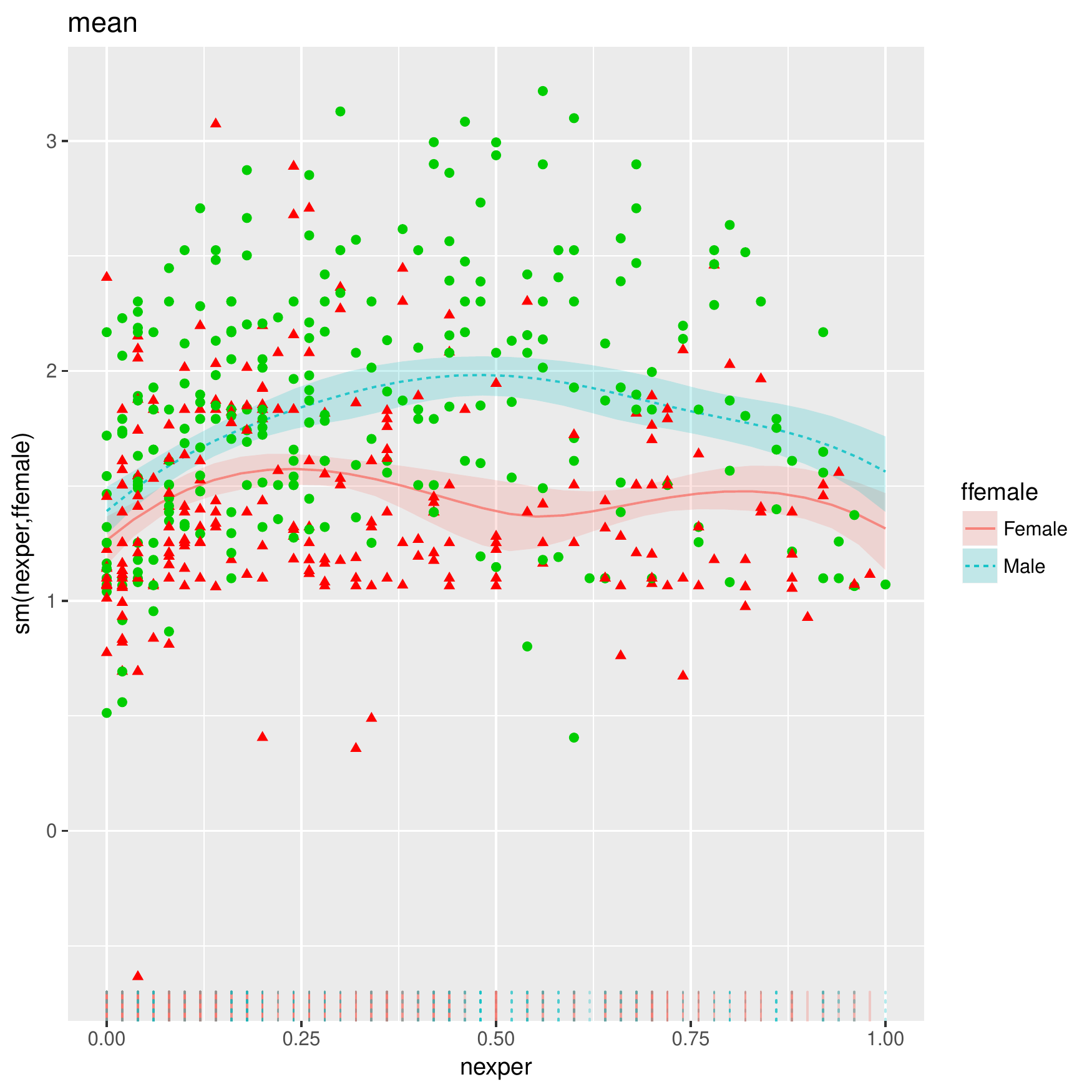} &  
\includegraphics[width=0.40\textwidth,height=0.16\textheight]{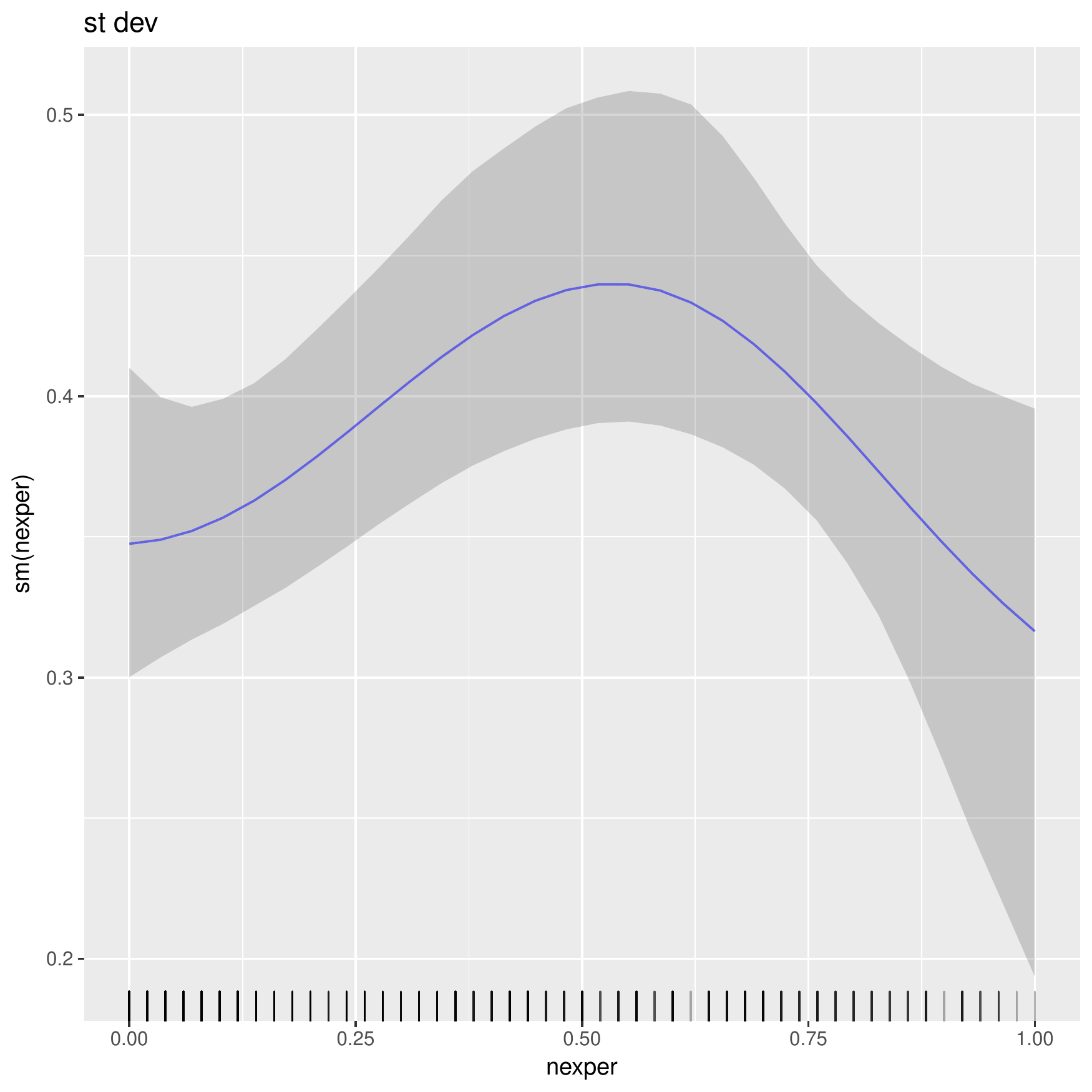} \\
(c) & (d) \\
\end{tabular}
\end{center}
\caption{Results from the data analysis on the relationship between covariates gender, marital status, experience, education 
and tenure, and response variable logarithm of hourly wage.
Posterior means and $95\%$ credible intervals for (a) $f_1(\texttt{ntenure})$, 
(b) $f_2(\texttt{neduc})$, (c) $f_3(\texttt{nexper},\texttt{female})$,
and (d) the standard deviation function $\sigma_i = \sigma \exp[f_4(\texttt{nexper})/2]$.}\label{da2}
\end{figure}

After summarizing results and checking convergence, we create plots of posterior means, along with  
$95\%$ credible intervals, for functions $f_1,\dots,f_4$. These are displayed in Figure~\ref{da2}.
As it turns out, variable \code{married} does not have an effect on the mean of \code{lwage}.
For this reason, we do not provide further results on the posterior of the coefficient of covariate \code{married}, $\beta_1$. 
However, in the code below we show how graphical summaries on $\beta_1$ can be obtained, if needed. 
\begin{verbatim}
> PlotOptionsT <- list(geom_point(data = wage1, aes(x = ntenure, y = lwage)))
> plot(x = m5, model = "mean", term="sm(ntenure)", quantiles = c(0.025, 0.975),
+      plotOptions = PlotOptionsT)
> PlotOptionsEdu <- list(geom_point(data = wage1, aes(x = neduc, y = lwage)))
> plot(x = m5, model = "mean", term = "sm(neduc)", quantiles = c(0.025, 0.975),
+      plotOptions = PlotOptionsEdu)
> pchs <- as.numeric(wage1$female)
> pchs[pchs == 1] <- 17; pchs[pchs == 2] <- 19
> cols <- as.numeric(wage1$female)
> cols[cols == 2] <- 3; cols[cols == 1] <- 2 
> PlotOptionsE <- list(geom_point(data = wage1, aes(x = nexper, y = lwage),
+                      col = cols, pch = pchs, group = wage1$female))
> plot(x = m5, model = "mean", term="sm(nexper,female)", quantiles = c(0.025, 0.975),
+      plotOptions = PlotOptionsE)
> plot(x = m5, model = "stdev", term = "sm(nexper)", quantiles = c(0.025, 0.975))
> PlotOptionsF <- list(geom_boxplot(fill = 2, color = 1))
> plot(x = m5, model = "mean", term = "married", quantiles = c(0.025, 0.975),
+      plotOptions = PlotOptionsF)
\end{verbatim}

Figure~\ref{da2}, panels (a) and (b) show the posterior means and $95\%$ credible intervals for $f_1(\texttt{ntenure})$ 
and $f_2(\texttt{neduc})$. It can be seen that expected wages increase with tenure and education, although there is 
high uncertainty over a large part of the range of both covariates. Panel (c) 
displays the posterior mean and a $95\%$ credible interval for $f_3$. We can see that although the forms
of the two functions are similar, i.e. the interaction term is not needed, males have higher expected wages than females. 
Lastly, panel (d) displays posterior summaries of the standard deviation function, $\sigma_i = \sigma \exp(f_4/2)$. 
It can be seen that variability first increases and then decreases as experience increases.

Lastly, we obtain predictions and credible intervals for the levels \code{"Married"} and \code{"Notmarried"}
of variable \code{fmaried} and  the levels \code{"Female"} and \code{"Male"} of variable \code{ffemale},
with variables \code{ntenure}, \code{nedc} and \code{nexper} fixed at their mid-range.
\begin{verbatim}
> p1 <- predict(m5, newdata = data.frame(fmarried = rep(c("Married", "Notmarried"), 2), 
+       ntenure = rep(0.5, 4), neduc = rep(0.5, 4), nexper = rep(0.5, 4), 
+       ffemale = rep(c("Female", "Male"), each = 2)), interval = "credible")

> p1
       fit      lwr      upr
1 1.321802 1.119508 1.506574
2 1.320400 1.119000 1.505272
3 1.913341 1.794035 2.036255
4 1.911939 1.791578 2.034832
\end{verbatim}
The predictions are suggestive of no `marriage' effect and of `gender' effect.

\subsubsection{Brain activity}

Here we analyse brain activity level data obtained by functional magnetic resonance imaging. The dataset is available 
in package \CRANpkg{gamair} \citep{gamSW} and it was previously analysed by \citet{Landou}. We are interested in three 
of the columns in the dataset. These are the response variable, \code{medFPQ}, which is calculated as the 
median over three measurements of `Fundamental Power Quotient' and the two covariates, \code{X} and \code{Y}, which 
show the location of each voxel.

The following R code loads the relevant data frame, removes two outliers and transforms the response variable, as was suggested by  \citet{gamSW}. 
In addition, it plots the brain activity data using function \code{levelplot} from package \CRANpkg{lattice} \citep{lats}. 
\begin{verbatim}
> data(brain)
> brain <- brain[brain$medFPQ > 5e-5, ]
> brain$medFPQ <- (brain$medFPQ) ^ 0.25
> levelplot(medFPQ ~ Y * X, data = brain, xlab = "Y", ylab = "X", 
+           col.regions = gray(10 : 100 / 100))
\end{verbatim}  

\begin{figure}
\begin{center}
\begin{tabular}{cc}
\includegraphics[width=0.40\textwidth]{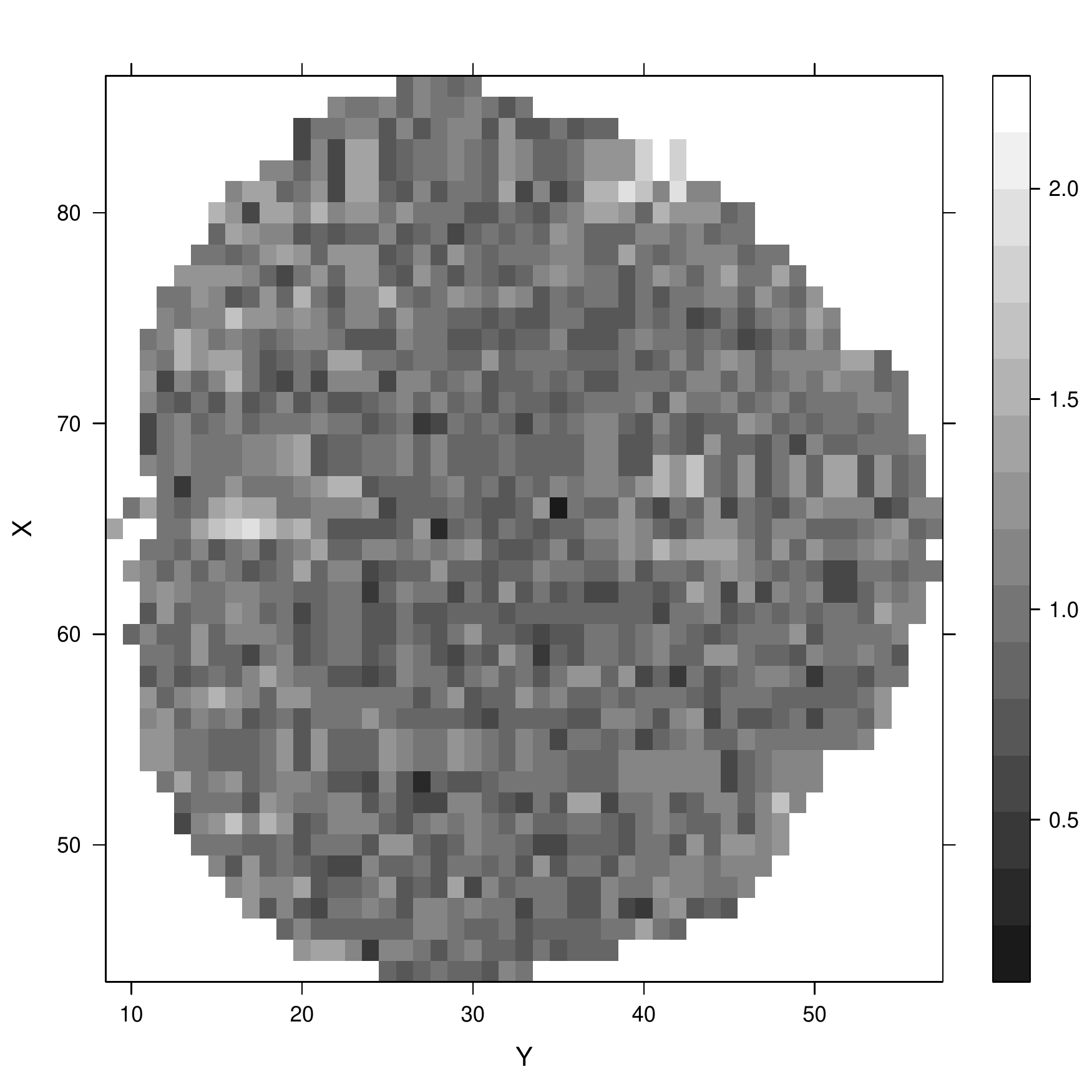} &  
\includegraphics[width=0.40\textwidth]{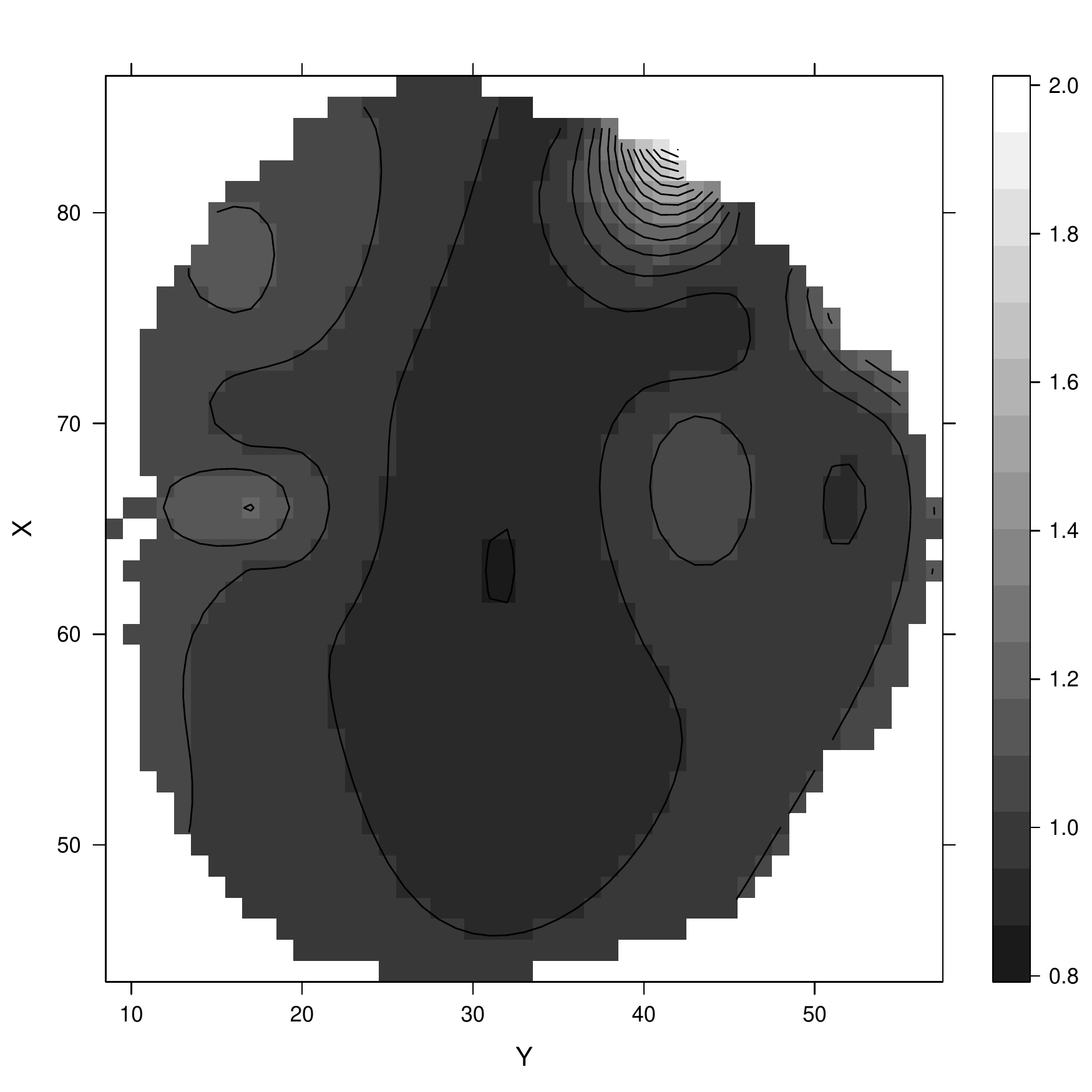} \\
(a) & (b) \\
\end{tabular}
\end{center}
\caption{Results from the brain activity level data analysis. Panel (a) shows the observed data and panel (b) the model-based smooth surface.}\label{brainact}
\end{figure}

The plot of the observed data is shown in Figure~\ref{brainact}, panel (a). Its distinctive feature is the noise level, which makes 
difficult to decipher any latent pattern. Hence, the goal of the current data analysis is to obtain a smooth surface of brain activity level from the noisy data. 
It was argued by \citet{gamSW} that for achieving this goal a spatial error term is not needed in the model. Thus, we analyse the brain activity level data using a model of the form
\begin{eqnarray}
\text{medFPQ}_i \stackrel{\text{ind}}{\sim} N(\mu_i,\sigma^2), \text{\;where\;} 
\mu_i = \beta_0 + \sum_{j_1=1}^{10} \sum_{j_2=1}^{10} \beta_{j_1,j_2} \phi_{1j_1,j_2}(X_i,Y_i), i=1,\dots,n, \nonumber
\end{eqnarray}
where $n=1565$ is the number of voxels. 

The R code that fits the above model is 
\begin{verbatim}
> Model <- medFPQ ~ sm(Y, X, k = 10, bs = "rd") | 1
> m6 <- mvrm(formula = Model, data = brain, sweeps = 50000, burn = 20000, thin = 2, 
+            seed = 1, StorageDir = DIR)
\end{verbatim}

From the fitted model we obtain a smooth brain activity level surface using function \code{predict}.
The function estimates the average activity at each voxel of the brain. Further, we plot the estimated surface 
using function \code{levelplot}.  
\begin{verbatim}
> p1 <- predict(m6)
> levelplot(p1[, 1] ~ Y * X, data = brain , xlab = "Y", ylab = "X", 
+           col.regions = gray(10 : 100 / 100), contour = TRUE)
\end{verbatim}

Results are shown in Figure~\ref{brainact}, panel(b). The smooth surface makes it much easier to see and understand which parts of the brain have higher  
activity. 

\subsubsection{Cars}

In the fourth and final application we use function \code{mvrm} to identify the best subset of predictors in a regression setting. 
Usually stepwise model selection is performed, using functions \code{step} and \code{stepAIC} in R. Here we show how \code{mvrm} 
can be used as alternative to those two functions. 
The data frame that we apply \code{mvrm} on is \code{mtcars}, where the response variable is \code{mpg}
and the explanatory variables that we consider are \code{disp}, \code{hp}, \code{wt} and \code{qsec}. The code below loads the data frame, 
specifies the model and obtains samples from the posteriors of the model parameters. 
\begin{verbatim}
> data(mtcars)
> Model <- mpg ~ disp + hp + wt + qsec | 1
> m7 <- mvrm(formula = Model, data = mtcars, sweeps = 50000, burn = 25000, thin = 2, 
+            seed = 1, StorageDir = DIR)
\end{verbatim}

The following is an excerpt of the output that function \code{summary} produces and it shows the three models with the highest posterior probability.
\begin{verbatim}
> summary(m7, nModels = 3)

Joint mean/variance model posterior probabilities:
mean.disp mean.hp mean.wt mean.qsec freq  prob cumulative
1       0       1       1         0 1085 43.40      43.40
2       0       0       1         1 1040 41.60      85.00
3       0       0       1         0  128  5.12      90.12
Displaying 3 models of the 11 visited
3 models account for 90.12% of the posterior mass
\end{verbatim}
The model with the highest posterior probability ($43.4\%$) is the one that includes explanatory variables \code{hp} and \code{wt}. 
The model that includes \code{wt} and \code{qsec} has almost equal posterior probability, $41.6\%$. 
These two models account of $85\%$ of the posterior mass. The third most promising model is the one that includes only \code{wt} 
as predictor, but its posterior probability is much lower, $5.12\%$. 

\section{Appendix: MCMC algorithm}

In this section we present the technical details of how the MCMC algorithm is designed for the case where 
there is a single covariate in the mean and variance models. We first
note that to improve mixing of the sampler, we integrate out vector $\ubeta$ from the likelihood 
of $\uy$, as was done by \citet{Chan06}
\begin{eqnarray}\label{marginal}
f(\uy|\ualpha,c_{\beta},\ugamma,\udelta,\sigma^2) \propto 
|\sigma^2 D^2(\ualpha_{\delta})|^{-\frac{1}{2}} (c_{\beta}+1)^{-\frac{N(\gamma)+1}{2}} \exp(-S/2\sigma^2),
\end{eqnarray}
where, with $\uty =  \uD^{-1}(\ualpha_{\delta}) \uy$, we have $S = S(\uy,\ualpha,c_{\beta},\ugamma,\udelta) = \uty^{\top} \uty - \frac{c_{\beta}}{1+c_{\beta}}\uty^{\top}\utX_{\gamma}
(\utX_{\gamma}^{\top}\utX_{\gamma})^{-1}\utX_{\gamma}^{\top}\uty$.

The six steps of the MCMC sampler are as follows
\begin{enumerate}

\item Similar to \citet{Chan06}, the elements of $\ugamma$ 
are updated in blocks of randomly chosen elements. The block size is chosen based on probabilities that can 
be supplied by the user or be left at their default values. Let $\ugamma_B$ be a block of random size of randomly chosen elements from $\ugamma$.
The proposed value for $\ugamma_B$ is obtained from its prior with the remaining elements 
of $\ugamma$, denoted by $\ugamma_{B^C}$, kept at their current value.   
The proposal pmf is obtained from the Bernoulli prior with $\pi_{\mu}$ integrated out
\begin{eqnarray}
p(\ugamma_{B}|\ugamma_{B^C}) = \frac{p(\ugamma)}{p(\ugamma_{B^C})}= 
\frac{\text{Beta}(c_{\mu}+N(\ugamma),d_{\mu}+q_1-N(\ugamma))}
{\text{Beta}(c_{\mu}+N(\ugamma_{B^C}),d_{\mu}+q_1-L(\ugamma_{B})-N(\ugamma_{B^C}))},\nonumber
\end{eqnarray}
where $L(\ugamma_{B})$ denotes the length of $\ugamma_{B}$ i.e. the size of the block. 
For this proposal pmf, the acceptance probability of the Metropolis-Hastings move reduces
to the ratio of the likelihoods in (\ref{marginal})
\begin{eqnarray}
\min\left\{1,\frac{(c_{\beta}+1)^{-\frac{N(\ugamma^P)+1}{2}} \exp(-S^P/2\sigma^2)}
{(c_{\beta}+1)^{-\frac{N(\ugamma^C)+1}{2}} \exp(-S^C/2\sigma^2)}\right\}, \nonumber
\end{eqnarray}
where superscripts $P$ and $C$ denote proposed and currents values respectively. 

\item Vectors $\ualpha$ and $\udelta$ are updated simultaneously.
Similarly to the updating of $\ugamma$, the elements of $\udelta$ are updated in random order in blocks 
of random size. Let $\udelta_B$ denote a block. Blocks $\udelta_B$ and the whole vector
$\ualpha$ are generated simultaneously. As was mentioned by \citet{Chan06}, generating the whole
vector $\ualpha$, instead of subvector $\ualpha_B$, is necessary
in order to make the proposed value of $\ualpha$ consistent with the proposed value of $\udelta$.  

Generating the proposed value for $\udelta_B$ is done in a similar way as was done for $\ugamma_B$ 
in the previous step. Let $\udelta^P$ denote the proposed value of $\udelta$. Next, we describe how  
the proposed vale for $\ualpha_{\delta^P}$ is obtained. The development is in the 
spirit of \citet{Chan06} who built on the work of \citet{Gamerman1997}. 

Let $\hat{\ubeta}_{\gamma}^C = \{c_{\beta}/(1+c_{\beta})\}(\utX_{\gamma}^{\top} \utX_{\gamma})^{-1}\utX_{\gamma}^{\top} \uty$
denote the current value of the posterior mean of $\ubeta_{\gamma}$.
Define the current squared residuals 
\begin{equation}
e_{i}^C = (y_{i} - (\ux^*_{i \gamma})^{\top} \hat{\ubeta}_{\gamma}^C)^2, \nonumber
\end{equation}
$i=1,\dots,n$.
These will have an approximate $\sigma^2_i \chi^2_1$ distribution,
where $\sigma^2_i = \sigma^2 \exp(\uz_i^{\top} \ualpha)$. 
The latter defines a Gamma generalized linear model (GLM) for the squared 
residuals with mean $E(\sigma^2_i \chi^2_1) = \sigma^2_i = \sigma^2 \exp(\uz_i^{\top} \ualpha)$, which,
utilizing a $\log$-link, can be thought of as Gamma GLM with an 
offset term: $\log(\sigma^2_i) = \log(\sigma^2) + \uz_i^{\top} \ualpha$.
Given the proposed value of $\udelta$, denoted by $\udelta^P$, the proposal density for $\ualpha^P_{\delta^P}$ 
is derived utilizing the one step iteratively re-weighted least squares algorithm.
This proceeds as follows. First define the transformed observations
\begin{eqnarray}
d_{i}^C(\ualpha^C) = \log(\sigma^2) + \uz_i^{\top} \ualpha^C + \frac{e_{i}^C-(\sigma^2_i)^C}{(\sigma^2_i)^C},\nonumber
\end{eqnarray}
where superscript $C$ denotes current values. Further, let $\ud^C$ denote the vector of $d_{i}^C$.

Next we define
\begin{eqnarray}
\uDelta(\udelta^P) = (c_{\alpha}^{-1}\uI + \uZ_{\delta^P}^{\top}\uZ_{\delta^P})^{-1}
\text{\;and\;}
\ahat(\udelta^P,\ualpha^C) =  \uDelta_{\delta^P} \uZ_{\delta^P}^{\top} \ud^C, \nonumber
\end{eqnarray}
where $\uZ$ is the design matrix.
The proposed value $\ualpha_{\delta^P}^P$ is obtained from a multivariate normal
distribution with mean $\ahat(\udelta^P,\ualpha^C)$ and covariance $h\uDelta(\udelta^P)$,
denoted as 
$N(\ualpha_{\delta^P}^P;\ahat(\udelta^P,\ualpha^C),h\uDelta(\udelta^P))$, where $h$ is a free parameter that 
we introduce and select its value adaptively \citep{roberts_examples_2009} in order to achieve an acceptance probability 
of $20\% - 25\%$ \citep{Roberts2001c}.

Let $N(\ualpha_{\delta^C}^C;\ahat(\udelta^C,\ualpha^P),h\uDelta(\udelta^C))$ denote the proposal density 
for taking a step in the reverse direction, from model $\udelta^P$ to $\udelta^C$. 
Then the acceptance probability of the pair $(\udelta^P,\ualpha^P_{\delta^P})$ is the minimum between 1 and  
\begin{eqnarray}
\frac{|D^2(\ualpha^P_{\delta^P})|^{-\frac{1}{2}} \exp(-S^P/2\sigma^2)}
{|D^2(\ualpha^C_{\delta^C})|^{-\frac{1}{2}} \exp(-S^C/2\sigma^2)}
\frac{
(2\pi c_{\alpha})^{-\frac{N(\delta^P)}{2}}\exp(-\frac{1}{2c_{\alpha}} (\ualpha^P_{\delta^P})^{\top} \ualpha^P_{\delta^P})
}{
(2\pi c_{\alpha})^{-\frac{N(\delta^C)}{2}}\exp(-\frac{1}{2c_{\alpha}} (\ualpha^C_{\delta^C})^{\top} \ualpha^C_{\delta^C})
}
\frac{
N(\ualpha_{\delta^C}^C;\ahat_{\delta^C},h\uDelta_{\delta^C})
}{
N(\ualpha_{\delta^P}^P;\ahat_{\delta^P},h\uDelta_{\delta^P})
}.\nonumber
\end{eqnarray}
We note that the determinants that appear in the above ratio are equal to one when utilizing centred explanatory variables in the 
variance model and hence can be left out of the calculation of the acceptance probability.  

\item We update $\sigma^2$ utilizing the marginal (\ref{marginal}) and the two priors in (\ref{PriorParams2}). 
The full conditional corresponding to the IG$(a_{\sigma},b_{\sigma})$ prior is
\begin{eqnarray}
f(\sigma^2|\dots) \propto (\sigma^2)^{-\frac{n}{2}-a_{\sigma}-1} \exp\{-(S/2+b_{\sigma})/\sigma^2\}, \nonumber
\end{eqnarray}
which is recognized as another inverse gamma IG$(n/2+a_{\sigma},S/2+b_{\sigma})$ distribution.

The full conditional corresponding to the normal prior  $|\sigma| \sim N(0,\phi^2_{\sigma})$ is
\begin{eqnarray}
f(\sigma^2|\dots) \propto (\sigma^2)^{-\frac{n}{2}} \exp(-S/2\sigma^2) \exp(-\sigma^2/2\phi^2_{\sigma}). \nonumber
\end{eqnarray}
Proposed values are obtained from $\sigma^2_p \sim N(\sigma^2_c,f^2)$ where $\sigma^2_c$ denotes the current value. 
Proposed values are accepted with probability $f(\sigma^2_p|\dots)/f(\sigma^2_c|\dots)$. We treat $f^2$ as a tuning 
parameter and we select its value adaptively \citep{roberts_examples_2009} in order to achieve an acceptance probability 
of $20\% - 25\%$ \citep{Roberts2001c}.

\item Parameter $c_{\beta}$ is updated from the marginal (\ref{marginal}) and the IG$(a_{\beta},b_{\beta})$ prior 
\begin{eqnarray}
f(c_{\beta}|\dots) \propto (c_{\beta}+1)^{-\frac{N(\gamma)+1}{2}} \exp(-S/2\sigma^2)
(c_{\beta})^{-a_{\beta}-1} \exp(-b_{\beta}/c_{\beta}). \nonumber
\end{eqnarray}
To sample from the above, we utilize a normal approximation to it. Let
$\ell(c_{\beta}) = \log\{f(c_{\beta}|\dots)\}$. We utilize a normal proposal density 
$N(\hat{c}_{\beta},-g^2/\ell^{''}(\hat c_{\beta})),$ where $\hat{c}_{\beta}$ is the mode
of $\ell(c_{\beta})$, found using a Newton-Raphson algorithm, 
$\ell^{''}(\hat c_{\beta})$ is the second derivative of $\ell(c_{\beta})$ evaluated at the mode, 
and $g^2$ is a tuning variance parameter the value of which is chosen adaptively \citep{roberts_examples_2009}. 
At iteration $u+1$ the acceptance probability is the minimum between one and
\begin{equation}
\frac{f(c_{\beta}^{(u+1)}|\dots)}{f(c_{\beta}^{(u)}|\dots)}
\frac{N(c_{\beta}^{(u)};\hat{c}_{\beta},-g^2/\ell^{''}(\hat c_{\beta}))}
{N(c_{\beta}^{(u+1)};\hat{c}_{\beta},-g^2/\ell^{''}(\hat c_{\beta}))}. \nonumber
\end{equation}

\item Parameter $c_{\alpha}$ is updated from the inverse Gamma density
IG$(a_{\alpha}+N(\delta)/2,b_{\alpha}+\ualpha_{\delta}^{\top}\ualpha_{\delta}/2)$.

\item The sampler utilizes the marginal in (\ref{marginal}) to improve mixing. 
However, if samples are required from the posterior of $\ubeta$, they can be generated from
\begin{eqnarray}
\ubeta_{\gamma} | \dots \sim N(\frac{c_{\beta}}{1+c_{\beta}}(\utX_{\gamma}^{\top} \utX_{\gamma})^{-1}\utX_{\gamma}^{\top} \uty,
\frac{\sigma^2 c_{\beta}}{1+c_{\beta}}(\utX_{\gamma}^{\top} \utX_{\gamma})^{-1}), \nonumber
\end{eqnarray}
where $\ubeta_{\gamma}$ is the non-zero part of $\ubeta$. 
\end{enumerate}

\section{Summary}

We have presented a tutorial on several functions from the R package \pkg{BNSP}. These  
functions are used for specifying, fitting and summarizing results from regression models with Gaussian 
errors and with mean and variance functions that can be modeled nonparametrically. 
Function \code{sm} is utilized to specify smooth terms in the mean and variance functions of the model.
Function \code{mvrm} calls an MCMC algorithm that obtains samples from the posteriors of the model
parameters. Samples are converted into an object of class `mcmc' by function \code{mvrm2mcmc} which facilitates 
the use of multiple functions from package \pkg{coda}. Functons \code{print.mvrm} and \code{summary.mvrm} provide
summaries of fitted `mvrm' objects. Further, function \code{plot.mvrm} provides graphical summaries of
parametric and nonparametric terms that enter the mean or variance function. Lastly, function \code{predict.mvrm}
provides predictions for a future response or a mean response along with the corresponding prediction/credible intervals. 

\bibliographystyle{apalike2}
\bibliography{all.bib}

\end{document}